\documentclass[a4paper,fleqn,usenatbib]{mnras}

\usepackage{newtxtext,newtxmath}

\usepackage[T1]{fontenc}
\usepackage{ae,aecompl}

\usepackage{graphicx}	
\usepackage{amsmath}	
\usepackage{amssymb}	

\title[Disappearing/emerging broad absorption lines]{Broad absorption line disappearance and emergence using multiple-epoch spectroscopy from the Sloan Digital Sky Survey}

\author[S. M. McGraw et al.]{S. M. McGraw,$^{1,2}$\thanks{E-mail: mcgrawsean01@gmail.com (SMM)}
W. N. Brandt,$^{1,2,3}$\thanks{E-mail: wnbrandt@gmail.com (WNB)}
C. J. Grier,$^{1,2}$
N. Filiz Ak,$^{4,5}$
P. B. Hall,$^{6}$ \newauthor
D. P. Schneider,$^{1,2}$
S. F. Anderson,$^{7}$
P. J. Green,$^{8}$
T. A. Hutchinson,$^{9}$ \newauthor
C. L. Macleod,$^{8}$ 
and M. Vivek$^{9}$
\\
$^{1}$Department of Astronomy \& Astrophysics, The Pennsylvania State University, 525 Davey Lab, University Park, PA 16802, USA\\
$^{2}$Institute for Gravitation and the Cosmos, The Pennsylvania State University, University Park, PA 16802, USA\\
$^{3}$Department of Physics, The Pennsylvania State University, University Park, PA 16802, USA\\
$^{4}$Faculty of Sciences, Department of Astronomy and Space Sciences, Erciyes University, 38039 Kayseri, TUR\\
$^{5}$Astronomy and Space Sciences Observatory and Research Center, Erciyes University, 38039 Kayseri, TUR\\
$^{6}$Department of Physics and Astronomy, York University, 4700 Keele Street, Toronto, Ontario M3J 1P3, CAN\\
$^{7}$Department of Astronomy, University of Washington, Box 351580, Seattle, WA 98195, USA\\
$^{8}$Harvard-Smithsonian Center for Astrophysics, 60 Garden Sreet, Cambridge, MA 02138, USA\\
$^{9}$Department of Physics and Astronomy, University of Utah, 115 S. 1400 E., Salt Lake City, UT 84112, USA\\
}

\date{Accepted XXX. Received YYY; in original form ZZZ}

\pubyear{2017}

\begin{document}
\label{firstpage}
\pagerange{\pageref{firstpage}--\pageref{lastpage}}
\maketitle

\begin{abstract}
We investigate broad absorption line (BAL) disappearance and emergence using a 470 BAL-quasar sample over $\le$ 0.10--5.25 rest-frame years with at least three spectroscopic epochs for each quasar from the Sloan Digital Sky Survey. We identify 14 disappearing BALs over $\le$ 1.73--4.62 rest-frame years and 18 emerging BALs over \mbox{$\le$ 1.46--3.66} rest-frame years associated with the C\thinspace\textsc{iv} $\lambda\lambda$1548,1550 and/or Si\thinspace\textsc{iv} $\lambda\lambda$1393,1402 doublets, and report on their variability behavior. BAL quasars in our dataset exhibit disappearing/emerging C\thinspace\textsc{iv} BALs at a rate of 2.3$^{+0.9}_{-0.7}$ and 3.0$^{+1.0}_{-0.8}$ per cent, respectively, and the frequency for BAL to non-BAL quasar transitions is 1.7$^{+0.8}_{-0.6}$ per cent. We detect four \emph{re-emerging} BALs over $\le$ 3.88 rest-frame years on average and three \emph{re-disappearing} BALs over $\le$ 4.15 rest-frame years on average, the first reported cases of these types. We infer BAL lifetimes along the line of sight to be nominally \mbox{$\la$ 100--1000}~yr using disappearing C\thinspace\textsc{iv} BALs in our sample. Interpretations of \mbox{(re-)emerging} and \mbox{(re-)disappearing} BALs reveal evidence that collectively supports both transverse-motion and ionization-change scenarios to explain BAL variations. We constrain a nominal C\thinspace\textsc{iv}/Si\thinspace\textsc{iv} BAL outflow location of $\la$ 100~pc from the central source and a radial size of $\ga$ 1$\times$10$^{-7}$~pc (0.02~au) using the ionization-change scenario, and constrain a nominal outflow location of $\la$ 0.5~pc and a transverse size of $\sim$0.01~pc using the transverse-motion scenario. Our findings are consistent with previous work, and provide evidence in support of BALs tracing compact flow geometries with small filling factors. 
\end{abstract}




\section{Introduction}

Outflows/winds are a significant component of the quasar phenomenon, and are plausible candidates for regulating potential co-evolution between the central supermassive black hole (SMBH) and the surrounding host galaxy (see e.g., \citealt{kor13}). Quasar winds might contribute toward establishing observed correlations between the SMBH mass and properties of the surrounding bulge, such as the \mbox{$M_{\bullet}$--$\sigma_*$} relation (\citealt{fer00}; \citealt{geb00}). Quasar outflows have also been invoked as possible candidates for active galactic nuclei (AGN) feedback processes, including termination of star formation in the interstellar medium (ISM) and regulation of accretion on to the SMBH (see e.g., \citealt{fab12}; \citealt{kin15}).  

Broad absorption lines (BALs) in rest-frame UV spectra are one of the primary tracers of quasar outflows. \citet{wey81} defined a BAL to be a trough that extends $\ge$~10 per cent below the continuum for $\ge$~2000~km~s$^{-1}$. BALs are generally associated with the C\thinspace\textsc{iv} $\lambda\lambda$1548,1550 and Si\thinspace\textsc{iv} $\lambda\lambda$1393,1402 doublets, and sources with these BALs in their spectra are referred to as high-ionization BAL (HiBAL) quasars. Low-ionization BAL (LoBAL) quasars additionally exhibit Al\thinspace\textsc{iii} $\lambda\lambda$1854,1862 and Mg\thinspace\textsc{ii} $\lambda\lambda$2796,2803 broad absorption in their spectra, while the rare iron low-ionization BAL (FeLoBAL) quasars additionally show broad absorption from Fe\thinspace\textsc{ii} or Fe\thinspace\textsc{iii} transitions. HiBALs are observed in an estimated 10--26 per cent of quasar spectra, while LoBALs and FeLoBALs are present in roughly 0.5--1 and $\la$~0.3 per cent of quasar spectra, respectively (see e.g., \citealt{hew03}; \citealt{tru06}; \citealt{kni08}; \citealt{gib09b}; \citealt{all11}).

Detecting BAL variability using multiple-epoch spectra is an important technique for constraining quasar wind lifetimes, locations, and geometries, which are critical for understanding the role of outflows in AGN feedback. Knowledge of these properties requires an understanding of the cause of BAL variations. BAL profiles can be altered either by a change in coverage of the background light source due to transverse motions of outflows across our line of sight (LOS), or by a change in optical depth within the absorbers due to fluctuations in the ionizing radiation field. \citet{cap13} discussed other possible scenarios for BAL variations, and reasons why they are unlikely. 

BAL variability studies have analysed a range of ions and time-scales, and have generally used small-to-intermediate sample sizes. For example, \citet{bar94} studied 23 quasars with C\thinspace\textsc{iv}, N\thinspace\textsc{v}, Si\thinspace\textsc{iv}, and Al\thinspace\textsc{iii} BALs over monthly to yearly rest-frame\footnote{All subsequent times are in the rest frame of the quasar.} time-scales. \citet{gib08} investigated C\thinspace\textsc{iv} BAL variability in 13 quasars over 3--6 yr. \citet{cap11,cap12,cap13} characterized C\thinspace\textsc{iv} and Si\thinspace\textsc{iv} BAL variations in 24 luminous quasars over 4--9~months and 3.8--7.7~yr. \citet{gri15} studied rapid C\thinspace\textsc{iv} BAL changes down to 1.20~d in the quasar SDSS J141007.74+541203.3. \citet{mcg15} analysed Fe\thinspace\textsc{ii} and Mg\thinspace\textsc{ii} BAL variability between 10~d and 7.6~yr in 12 FeLoBAL quasars. \citet{ham08} discovered the first example of emergence of a high-velocity C\thinspace\textsc{iv}/Si\thinspace\textsc{iv} broad-line outflow over $\sim$1.54~yr in the quasar J105400.40+034801.2. \citet{hal11} observed the disappearance of Fe\thinspace\textsc{ii} troughs and dramatic variability of the Mg\thinspace\textsc{ii} BAL at the same velocity over 0.6--5~yr in the FeLoBAL quasar J1408+3054.

The Sloan Digital Sky Survey (SDSS) has successfully enabled large-scale investigations of BAL variability over multi-year time-scales. \citet{fil12} conducted a systematic study of 582 bright BAL quasars over \mbox{1.1--3.9}~yr, and detected 21 cases of C\thinspace\textsc{iv} BAL disappearance. \citet{fil13} carried out a variability analysis of C\thinspace\textsc{iv} and Si\thinspace\textsc{iv} BALs over \mbox{1--3.7}~yr using a sample of 291 BAL quasars. \citet{fil14} studied the correlated behavior between C\thinspace\textsc{iv}, Si\thinspace\textsc{iv}, and Al\thinspace\textsc{iii} BALs over multi-year time-scales using a sample of 671 BAL quasars. \citet{gri16} performed a variability study of 140 BAL quasars over \mbox{2.5--5.5}~yr using three spectroscopic epochs, and measured upper limits for C\thinspace\textsc{iv} BAL acceleration and deceleration in 76 troughs to find that most BALs in their sample exhibit stable mean velocities to within about 3 per cent.

In this work we use quantitative criteria to conduct a systematic analysis of C\thinspace\textsc{iv} and Si\thinspace\textsc{iv} BAL disappearance and emergence using a sample of 470 BAL quasars over \mbox{0.10--5.25}~yr in the rest frame. We build upon the BAL-disappearance work of \citet{fil12} by utilizing three or more spectroscopic epochs that are well separated in time for each BAL quasar, allowing us to investigate the behavior during and after BAL disappearance and emergence. We present here a number of cases of BAL \emph{re-emergence} and \emph{re-disappearance} events, which to our knowledge are the first reported cases of these types.\footnote{\citet{fil12} reported the first example of a C\thinspace\textsc{iv} BAL re-disapperanace event (see their fig.~5), and we discuss this case in Section 4.3.} Our study also marks the first investigation to our knowledge into the relative behavior of disappearing and emerging BALs using a single sample of BAL quasars. We emphasize that our sample consists only of BAL quasars; investigations into the emergence of BAL outflows using a non-BAL quasar sample is beyond the scope of this work.

Section 2 of this paper outlines the spectroscopic observations and procedures used for constructing the sample of 470 BAL quasars. Section 3 details our analytical approach for fitting the continuum and our statistical criteria for detecting disappearing and emerging BALs. Section 4 presents results on the properties of disappearing and emerging BALs, re-emerging and re-disappearing BALs, and their quasar hosts. Section 5 discusses the origin of BAL disappearance and emergence, and places order-of-magnitude estimates on BAL outflow lifetimes, locations, and sizes. Section 6 presents the relevant conclusions of this study and outlines future work. Throughout this paper, we adopt a cosmology with $\Omega_{\Lambda}=0.7$, $\Omega_{\rm{M}}=0.3$, and $h=0.7$. All reported time-scales are in the rest frame of the quasar.

\section{Observations and Sample Selection}

The sample of BAL quasars used in this study has spectroscopic observations from \mbox{SDSS-I/II} (hereafter SDSS; \citealt{yor00}), the SDSS-III Baryon Oscillation Spectroscopic Survey (BOSS; \citealt{eis11}; \citealt{daw13}), and the SDSS-IV \citep{bla17} Time Domain Spectroscopic Survey (TDSS; \citealt{mor15}). Spectra from SDSS (\citealt{gun06}; \citealt{sme13}) cover wavelengths between \mbox{$\sim$3800--9200}~\AA, have pixels 70~km~s$^{-1}$ wide, and have resolutions ranging from \mbox{$\sim$1850} to 2200. BOSS and TDSS spectra were/are acquired using a different spectrograph, and cover wavelengths from \mbox{$\sim$3600} to 10\,000~\AA, have pixels 70~km~s$^{-1}$ wide, and have resolutions between \mbox{$\sim$1300--3000}. \citet{fil12,fil13} and \citet{gri16} provide more information regarding the above surveys and their ancillary programs that are relevant to this work, which specifically focus on conducting large-sample BAL-variability studies over multi-year time-scales.

Our initial sample consists of 2005 BAL quasars that are a subset of the 5039 BAL quasar sample constructed by \citet{gib09b} from the SDSS Data Release 5 (DR5) quasar catalog \citep{sch07}. The 2005 objects were chosen to be optically bright ($i$~$<$~19.3), have SDSS spectra with high signal-to-noise ratios (SNR$_{1700}$~$>$~6 as defined by \citealt{gib09b}), have spectroscopic coverage of wavelength intervals where the C\thinspace\textsc{iv}, Si\thinspace\textsc{iv}, Al\thinspace\textsc{iii}, and Mg\thinspace\textsc{ii} BALs are expected to arise, and exhibit relatively strong BALs in their spectra (see \citealt{gib09b}). 

We searched for available spectra by matching the coordinates of the 2005 BAL quasars to source positions from the SDSS, BOSS, and TDSS catalogs using a tolerance of 1.5~arcsec. The search retrieved 4814 spectra of the 2005 quasars: 2005 quasars with 2373 spectra from SDSS, 1606 quasars with 1981 spectra from BOSS, and 440 quasars with 460 spectra from TDSS. There are 388 quasars that have at least one available epoch from each of the three surveys (SDSS, BOSS, and TDSS). The observation dates for the retrieved spectra range from MJDs 51578 to 55004 (2000~February~4 to 2009~June~22) for SDSS, 55176 to 56837 (2009~December~11 to 2014~June~29) for BOSS, and 56900 to 57544 (2014~August~31 to 2016~June~5) for TDSS.

In order to conduct a systematic investigation of the behavior during and after BAL disappearance and emergence, we established a number of selection criteria to construct our sample starting from the 2005 BAL quasars with 4814 spectra. We only consider quasars with at least three available epochs that are well separated in time (i.e., each pair of spectra taken greater than 0.1~yr apart); 578 quasars with 1856 spectra met this criterion. Our analysis focuses on the C\thinspace\textsc{iv} $\lambda\lambda$1548,1550 and Si\thinspace\textsc{iv} $\lambda\lambda$1393,1402 BALs, and we therefore only consider quasars with redshifts of 1.68~$<$~$z$~$<$~4.93. These redshift limits yielded 548 quasars with 1760 spectra. We require that all spectra should be of high-quality (i.e., with SNR$_{1700}$~$>$~6 as defined by \citealt{gib09b}); 532 quasars with 1706 spectra passed this additional test. We also consider only BAL troughs that are significantly detached from nearby broad emission lines (BELs) (i.e., the entire BAL trough must lie between \mbox{--3000} and \mbox{--27\,600}~km~s$^{-1}$ from the C\thinspace\textsc{iv} BEL or between \mbox{--2000} and \mbox{--20\,800}~km~s$^{-1}$ from the Si\thinspace\textsc{iv} BEL).\footnote{We adopt a smaller limit of \mbox{--2000}~km~s$^{-1}$ since the Si\thinspace\textsc{iv} BEL is expected to be weaker than the C\thinspace\textsc{iv} BEL (see e.g., fig. 6 from \citealt{van01}).} The above criteria produce a final sample of 470 BAL quasars with 1509 spectra.

In addition to analysing BAL disappearance and emergence using spectroscopic observations, we utilize photometric observations from the Catalina Sky Survey (CSS) data release 2 \citep{dra09}. CSS synthesizes $V$-band fluxes using observations taken with an unfiltered CCD, and we use these measurements to supplement our discussion regarding the origin of BAL disappearance and emergence (see Section 5.1.2). In Section 5.1.2 we present synthesized $V$-band light curves which provide evidence for variations in continuum radiation that might be an indicator of fluctuations in the ionizing radiation field (see Fig.~\ref{fig:css}); this behavior provides evidence supporting the ionization-change scenario to explain BAL disappearance and emergence.

Our final 470 BAL-quasar sample is different from the samples constructed by \citet{fil12,fil13}, which were selected from the same initial sample of 2005 BAL quasars (see above). The sample from \citet{fil12} utilized observations from SDSS and BOSS prior to 2011 September 7, and the sample from \citet{fil13} made use of spectra from SDSS and BOSS prior to 2012 July 1; our sample makes use of SDSS, all of BOSS (i.e., through 2014 June 22), and part of TDSS. We also impose different limits for BAL velocities and require that spectra are separated by $>$0.1\,yr, while \citet{fil12,fil13} focused on time-scales $>$1\,yr. \citet{fil13} also enforced higher signal-to-noise and redshift cuts than the ones enforced in this study.

\section{Analysis}

\subsection{Continuum fits}

We performed a number of operations on the 1509 spectra and their associated statistical error arrays prior to conducting continuum fits. The BOSS spectra do not have reliable absolute flux calibrations, and we applied flux corrections when available (i.e., for 44 per cent of BOSS spectra) using the estimates from \citet{mar16}. For each spectrum we utilized the BRIGHTSKY bit mask to flag and interpolate pixels associated with insufficient night-sky subtraction. Spectra were corrected for Galactic extinction using the reddening curve from \citet{car89} and the $A_{\textrm{v}}$ values from \citet{sch11} assuming $R_{\textrm{v}}$~$=$~3.1. We utilize the redshift estimates from \citet{hew10} and converted spectra into the rest frame of each quasar.

\begin{figure*}
	\centering
	\includegraphics[width=1.75\columnwidth]{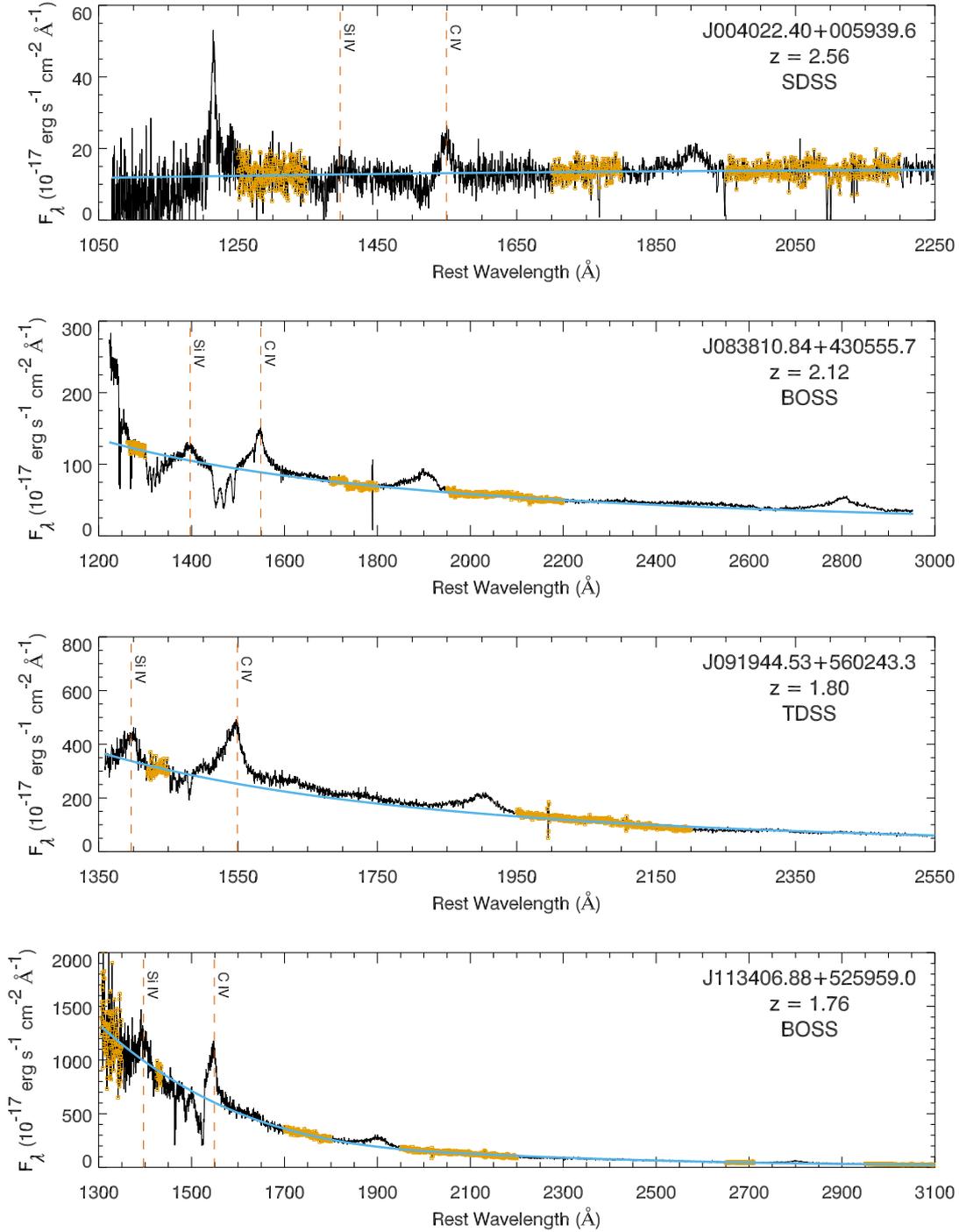}
	\caption{A series of representative BAL quasar spectra (black) and their associated continuum fits (solid, light blue lines). The quasar name, redshift, and survey is shown in the top-right corner of each plot. The orange data intervals represent the points within the chosen RLF windows that were not rejected by the sigma-clipping algorithm (see Section 3.1). The labelled, vertical, dashed red lines show the rest frame wavelengths of the C\thinspace\textsc{iv} and Si\thinspace\textsc{iv} BELs.}
	\label{fig:con}
\end{figure*}

The continuum in each of the 1509 spectra was initially modeled using an SMC-like reddened power-law function from \citet{pei92}, and free parameters include the power-law normalization, power-law index, and extinction coefficient. Following \citet{gib09b}, we initially applied the continuum function to relatively line-free (RLF) windows when available, which range between \mbox{1250--1350}, \mbox{1700--1800}, \mbox{1950--2200}, \mbox{2650--2710}, and \mbox{2950--3700}~\AA. For each spectrum we employed an iterative sigma-clipping algorithm that consists of fitting the continuum function to the RLF windows using a Levenberg-Marquardt least-squares minimization procedure \citep{mar09}, rejecting data points that deviate more than 3$\sigma$ from the fit within the RLF windows,\footnote{The $\sigma$ represents the standard deviation of the differences between the data points and the fit within the RLF windows.} and repeating these steps until the optimal fit is unchanged. The continuum fits were limited by the SDSS wavelength coverage for all epochs in each source. Fig.~\ref{fig:con} presents representative examples of spectra and their associated continuum fits.

We determined if each continuum fit was reasonable by visually inspecting the available RLF windows and the region between \mbox{1300--1900}~\AA \ where C\thinspace\textsc{iv}, Si\thinspace\textsc{iv}, and Al\thinspace\textsc{iii} BALs are expected to be detected. Continuum fits for \mbox{72} per cent of spectra employed the sigma-clipping algorithm and were applied to all available RLF windows listed above. The remaining spectra required additional actions in order to obtain a suitable fit, including removing the sigma-clipping algorithm to prevent useful RLF windows from being `clipped away', using a subset of the RLF windows, using smaller intervals within the RLF windows, including an additional interval within the range \mbox{1420--1510}~\AA \ that appeared free of absorption,\footnote{\citet{van01} demonstrated that this wavelength interval is also mostly free of emission (see their fig. 6).} and/or using a cubic spline function instead of a reddened power law. These additional actions were necessary due to the presence of broad, blended emission or strong absorption within some RLF windows, and/or unphysical continuum curves due to incorrect flux calibrations. Of the remaining epochs that did not yield successful fits using the sigma clipping and standard RLF windows listed above, \mbox{15} per cent required removal of the sigma-clipping algorithm and/or altering the RLF windows (see e.g., \mbox{J091944.53+560243.3} in Fig.~\ref{fig:con}) while \mbox{13} per cent additionally required the use of a cubic spline function of varying order between 1 and 5 (see e.g., \mbox{J113406.88+525959.0} in Fig.~\ref{fig:con}). Given that we do not relate continuum-fit parameters to any physical meaning, the use of alternative techniques for this small fraction of spectra does not cause loss of information for further analysis.

\subsection{Detection of disappearing and emerging BALs}

The 1509 continuum-normalized spectra were examined to identify C\thinspace\textsc{iv} and Si\thinspace\textsc{iv} BALs that are detached from their respective BELs (i.e., between \mbox{--3000} and \mbox{--27\,600}~km~s$^{-1}$ from the C\thinspace\textsc{iv} BEL or between \mbox{--2000} and \mbox{--20\,800}~km~s$^{-1}$ from the Si\thinspace\textsc{iv} BEL). We also search for Al\thinspace\textsc{iii} $\lambda\lambda$1854,1862 BALs within the same velocity limits as for C\thinspace\textsc{iv} to characterize the number of LoBAL quasars in our sample.\footnote{LoBAL quasars are completely identified by searching for both Al\thinspace\textsc{iii} and Mg\thinspace\textsc{ii} $\lambda\lambda$2796,2803 BALs--however, our current dataset does not allow for a significant number of Mg\thinspace\textsc{ii} BAL detections primarily due to wavelength coverage limitations. Thus our LoBAL classification should not be taken to be complete.} Prior to searching for BALs, we smoothed all spectra using a 5-pixel wide boxcar filter ($\sim$2 times the resolution of the SDSS spectrographs) to reduce pixel-to-pixel noise. Smoothed spectra were used for detecting BALs and estimating BAL velocities and absorption strengths (see below). We flagged troughs that were consistently $\ge$~10 per cent below the continuum level for $\ge$~2000~km~s$^{-1}$ based on the traditional definition of a BAL (Weymann et al. 1981). We require that the entire BAL trough exist within the velocity limits noted above to be included in our final sample.  

We detect 2228 BALs within the specified velocity intervals using the 1509 spectra associated with the 470 BAL quasars. For each detected BAL we measure a BAL velocity by calculating the mean of the minimum and maximum velocities of the BAL. We also determine the strength of each BAL by measuring its balnicity index BI and rest-frame equivalent width EW
\begin{equation}
\textrm{BI} \equiv \int^{v_{\rm{max}}}_{v_{\rm{min}}} \bigg[1-\frac{f(v)}{0.9}\bigg] C \ dv
\end{equation} 
\begin{equation}
\textrm{EW} \equiv \int^{\lambda_{\rm{max}}}_{\lambda_{\rm{min}}} \Big[1-f(\lambda)\Big] \ d\lambda
\end{equation}
In equation (1), $f(v)$ is the normalized flux density at some velocity $v$ and $C$ is a constant that is one when a trough is $\ge$~10 per cent below the continuum for $\ge$~2000~km~s$^{-1}$ and zero otherwise. In equation (2), $f(\lambda)$ is the normalized flux density at some rest-frame wavelength $\lambda$. The integration limits range from the minimum velocity $v_{\rm{min}}$ (or minimum rest-frame wavelength $\lambda_{\rm{min}}$) to maximum velocity $v_{\rm{max}}$ (or maximum rest-frame wavelength $\lambda_{\rm{max}}$) for each individual BAL.\footnote{Our calculation of the BI parameter for each individual BAL in a single spectrum differs from traditional methods utilized by previous work, which involved calculating a single BI value for all BALs of the same transition in a single spectrum.} Measurements for the first 10 of 2228 detected BALs are listed in Table~\ref{tab:cat}, and the full table is available in the online, supplementary material (see Table 1A).

\begin{table*}
\centering
\caption{BAL measurements and associated observations}
\label{tab:cat}
\begin{tabular}{ccrrrr}
\hline
BAL quasar & SDSS specID & $v_{\rm{min}}$ & $v_{\rm{max}}$ & BI & EW \\
& (plate-mjd-fiber) & (km s$^{-1}$) & (km s$^{-1}$) & (km s$^{-1}$) & (\AA) \\
\hline

J000912.89+000504.4 & 1491-52996-0355 & --22\,626 & --17\,667 & 193 &  4.0 \\
J000912.89+000504.4 & 4217-55478-0900 & --22\,945 & --18\,511 & 150 & 3.4 \\
J000912.89+000504.4 & 7862-56984-0148 & --22\,115 & --14\,598 & 295 & 5.7 \\
J001025.90+005447.6 & 0389-51795-0332 & --40\,926 & --35\,201 & 1578 & 13.5 \\
J001025.90+005447.6 & 0389-51795-0332 & --17\,189 & --13\,654 & 438 & 5.9 \\ 
J001025.90+005447.6 & 0389-51795-0332 & --12\,531 & --7797 & 1356 & 13.7 \\
J001025.90+005447.6 & 4218-55479-0592 & --40\,926 & --35\,201 & 1263 & 11.4 \\
J001025.90+005447.6 & 4218-55479-0592 & --24\,511 & --20\,296 & 288 & 4.2 \\
J001025.90+005447.6 & 4218-55479-0592 & --20\,038 & --15\,165 & 302 & 5.8 \\
J001025.90+005447.6 & 4218-55479-0592 & --14\,838 & --7594 & 3153 & 19.9 \\
\bf{.} & \bf{.} & \bf{.} & \bf{.} & \bf{.} & \bf{.} \\
\bf{.} & \bf{.} & \bf{.} & \bf{.} & \bf{.} & \bf{.} \\
\bf{.} & \bf{.} & \bf{.} & \bf{.} & \bf{.} & \bf{.} \\
\hline
 \end{tabular} \\
 \emph{Note.} Columns from left to right include: BAL quasar name, plate-mjd-fiber number for the associated SDSS spectrum, minimum BAL velocity, maximum BAL velocity, balnicity index, and rest-frame equivalent width.  BAL velocities are calculated with respect to the average wavelength of the C\thinspace\textsc{iv} doublet (i.e., 1549.062\,\AA). The full table is available in the online, supplementary material (see Table~1A).
\end{table*}

In order to characterize statistically BAL disappearance and emergence in our sample, we applied quantitative tests to pairwise comparisons between spectra for each quasar. For each comparison between two spectra $S_1$ and $S_2$ observed at times $t_1$ and $t_2$, we established two criteria to determine whether a BAL disappearance\footnote{The criteria are described in terms of finding a BAL disappearance event. The criteria for finding BAL emergence events are the same, but with the roles of $S_1$ and $S_2$ switched.} event occurred, and a third criterion to determine whether the variability was significant:\\ 

1. If a BAL is detected in $S_1$, then $S_2$ must not exhibit a BAL or mini-BAL over the wavelength interval where the BAL in $S_1$ is detected. We define a mini-BAL as a trough that is consistently $\ge$~10 per cent below the continuum over a \mbox{500--2000}~km~s$^{-1}$ range (see e.g., \citealt{ham04}). The presence of narrow absorption lines (NALs) with line widths \mbox{$<500$}~km~s$^{-1}$ could potentially lead to failed detections of disappearing BALs, and we address this issue at the end of the section.\\

2. The average flux $F_2$ in $S_2$ over the wavelength interval where the BAL is present in $S_1$ must be greater than the continuum level minus 4 times $\sigma_2$. In other words, the flux within the specified interval must be consistent with a flat continuum to within $-4\sigma_2$. The quantity $\sigma_2$ is the propagated, statistical error associated with $F_2$, and both are calculated using the following equations
\begin{equation}
F_2=\frac{\sum\limits_{i=1}^{N} \ f_i}{N} \ \ \ , \ \ \ 
\sigma_2=\frac{\sqrt{\sum\limits_{i=1}^{N} \ \sigma_i^2}}{N}
\end{equation} 
where $f_i$ and $\sigma_i$ are the flux and statistical error at each pixel $i$, and $N$ is the number of pixels associated with the BAL in $S_1$. The threshold of 4$\sigma_2$ is a conservative estimate designed to account for uncertainties associated with fitting the continuum and other systematic errors. The 4$\sigma_2$ threshold was determined by visual inspection to confirm that little residual absorption exists for comparisons meeting this condition.\\

3. The average flux difference $F_{2-1}$ between $S_2$ and $S_1$ over the wavelength interval where the BAL is present in $S_1$ must be greater than 7 times $\sigma_{2-1}$. This means that the flux within the specified interval must change more than 7$\sigma_{2-1}$ between the two spectra. The quantity $\sigma_{2-1}$ is the propagated, statistical error associated with $F_{2-1}$, and both are calculated using the following equations
\begin{equation}
F_{2-1}=F_2-F_1 \ \ \ , \ \ \ 
\sigma_{2-1}=\sqrt{\sigma_2^2+\sigma_1^2}
\end{equation} 
where $F_2$ and $\sigma_2$ are defined in equation (3), and $F_1$ and $\sigma_1$ are defined similarly. The 7$\sigma_{2-1}$ threshold is a conservative estimate designed to account for systematic uncertainties, and was determined by visual inspection to confirm that BAL variations meeting this condition exhibit large, convincing flux differences. \\

\begin{figure*}
\includegraphics[width=2.00\columnwidth]{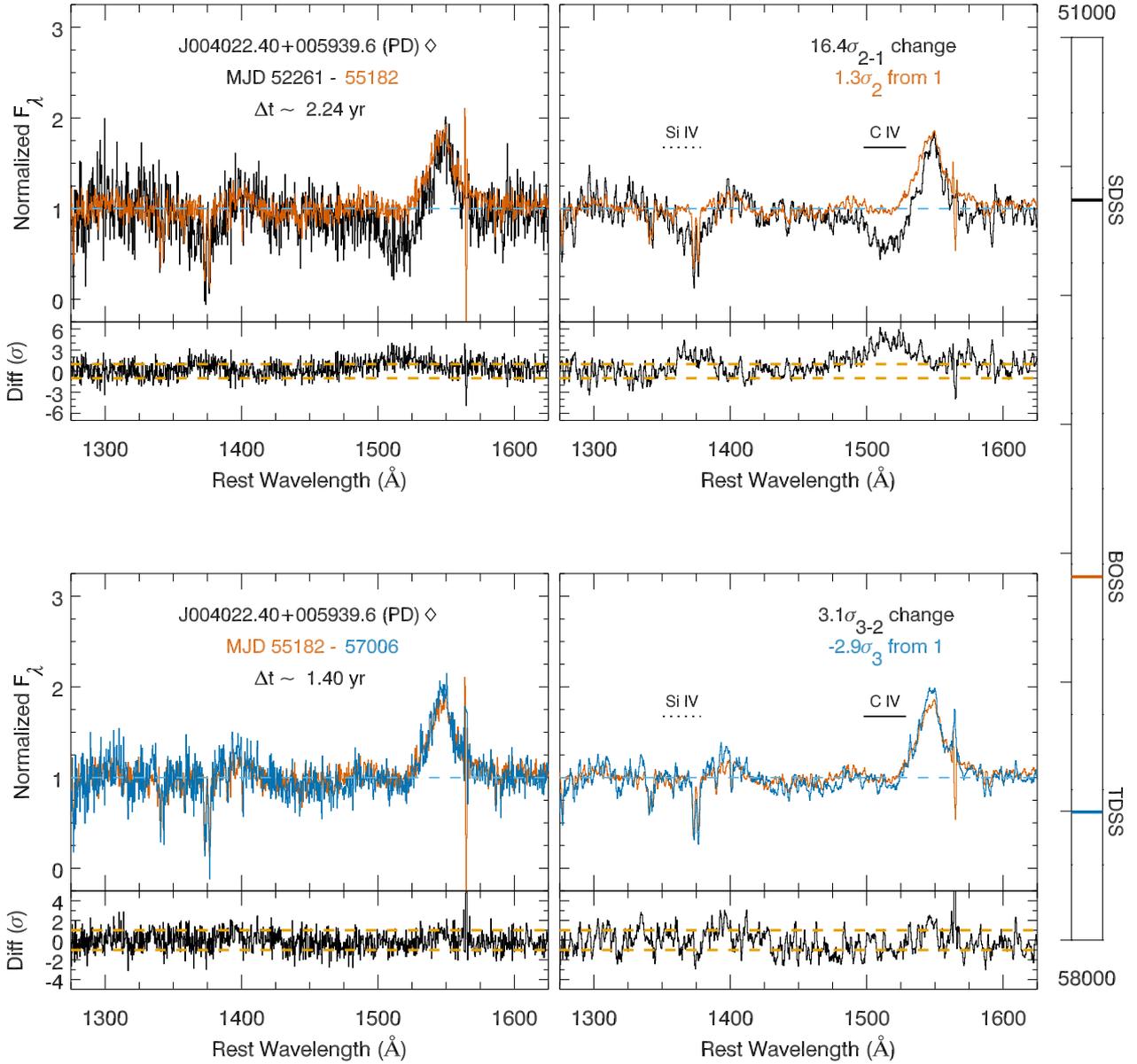}
\caption{Normalized spectrum comparisons showing 32 pristine cases of disappearing or emerging BALs. Each horizontal pair of panels displays the same comparison and plots un-smoothed spectra (left) and smoothed spectra using a 5-pixel boxcar (right). Each BAL quasar has two rows of panels; the first row compares spectrum one (black) and two (red) and reveals the BAL disappearance or emergence event; the second row compares spectrum two (red) and three (blue) and shows the behavior after the BAL disappears or emerges. Each spectrum comparison plot is accompanied by a plot directly beneath it which shows the flux difference (in units of $\sigma$) as well as $\pm1\sigma$ values (horizontal, dashed orange lines). Flux differences are calculated following \citet{fil12} [see their equation (2)]. We note that errors associated with smoothed flux differences are not statistically independent of one another due to the 5-pixel boxcar filter. For each quasar there is a narrow vertical panel on the right that plots time from MJD 51000 (top) to 58000 (bottom) and displays the MJD position, colour, and survey for each spectrum (tick marks are incremented by 1000~d). The dashed, horizontal light blue line at 1.0 in all panels indicates the continuum level.  Each left panel lists the quasar name, whether the quasar exhibits a pristine case of BAL disappearance (PD) or emergence (PE), observation dates and colours used for plotting the spectra, and the time-scale between epochs. A black-filled diamond next to a quasar name indicates the presence of a re-emerging or re-disappearing BAL (see Table~\ref{tab:pro} and Sections~4.2 and 4.3), and a white-filled diamond indicates sources that undergo a BAL to non-BAL quasar transition (see Table~\ref{tab:pro} and Section~4.1). The example panel shown above corresponds to a ``white-filled diamond" case. Labelled, horizontal solid lines in the right panels represent the wavelength intervals where disappearing or emerging BALs are detected from C\thinspace\textsc{iv} and/or Si\thinspace\textsc{iv}. Labelled, horizontal dotted lines lie at the same velocity as the disappearing or emerging BALs but correspond to the other ion. The first-line statement in each right panel is the significance of variability for the BAL change, and each factor of sigma displayed is calculated using $F_{2-1}/\sigma_{2-1}$ (see criterion~3). The second-line statement in each right panel is the consistency between the continuum and spectrum of the same colour as the text, and each factor of sigma displayed is calculated using $(F_{2}-1.0)/\sigma_{2}$ (see criterion~2). All the component figures are available in the online, supplementary material (see Fig.~2A).}
\label{fig:var}
\end{figure*}

The BAL variations that meet conditions 1, 2, and 3 listed above are considered to be part of a `pristine' sample of BAL disappearance and emergence events. The pristine sample is made up of cases where the  BAL variability is highly significant and there is little residual absorption after the BAL disappears or before the BAL emerges. Fig.~\ref{fig:var} shows BAL disappearance and emergence events in the pristine sample along with later observations showing the behavior after the BAL disappears or emerges.\footnote{Fig.~\ref{fig:var} displays the first component figure from Fig.~2A, which is available in the online, supplementary material (we only refer to Fig.~\ref{fig:var} hereafter). Fig.~\ref{fig:var} presents 29 component figures of 29 quasars each with a single, pristine case of BAL disappearance or emergence, displays two figures of the quasar J130542.35+463529.9 which show a pristine case of BAL emergence and a pristine case of BAL disappearance, and presents one figure of the quasar J081102.91+500724.2 which shows two pristine cases of BAL disappearance.}

Our pristine sample only contains the convincing cases of BAL disappearance and emergence upon visual inspection. In order to establish that the pristine sample is a near-complete set of BAL disappearance and emergence events given the quality of our data, we revise the criteria above to generate a `supplemental' sample using less-conservative thresholds. The revised criteria for supplemental sources are listed below in the case of a BAL disappearance event:\\

A. If a BAL is detected in $S_1$, then $S_2$ must not exhibit a BAL or mini-BAL over the wavelength interval where the BAL in $S_1$ is detected (same as criterion 1). We address the presence of NALs near disappearing BALs at the end of the section.\\

B. The average flux $F_2$ in $S_2$ over the wavelength interval where the BAL is present in $S_1$ must be between the continuum minus 7$\sigma_2$ and the continuum minus 4$\sigma_2$. This criterion allows for more residual absorption than criterion 2, as the flux over the specified interval must be consistent with a flat continuum to within $-7\sigma_2$ but not within $-4\sigma_2$. The quantity $\sigma_2$ is the propagated, statistical error associated with $F_2$, and both are calculated as shown in equation (3).\\

C. The average flux difference $F_{2-1}$ between $S_2$ and $S_1$ over the wavelength interval where the BAL is present in $S_1$ must be between 5$\sigma_{2-1}$ and 7$\sigma_{2-1}$. This criterion allows for less significant flux changes than criterion 3, as the flux within the specified interval must change more than 5$\sigma_{2-1}$ but not more than 7$\sigma_{2-1}$ between the two spectra. The quantity $\sigma_{2-1}$ is the propagated, statistical error associated with $F_{2-1}$, and both are calculated as shown in equation (4).\\

\begin{figure*}
\includegraphics[width=2.00\columnwidth]{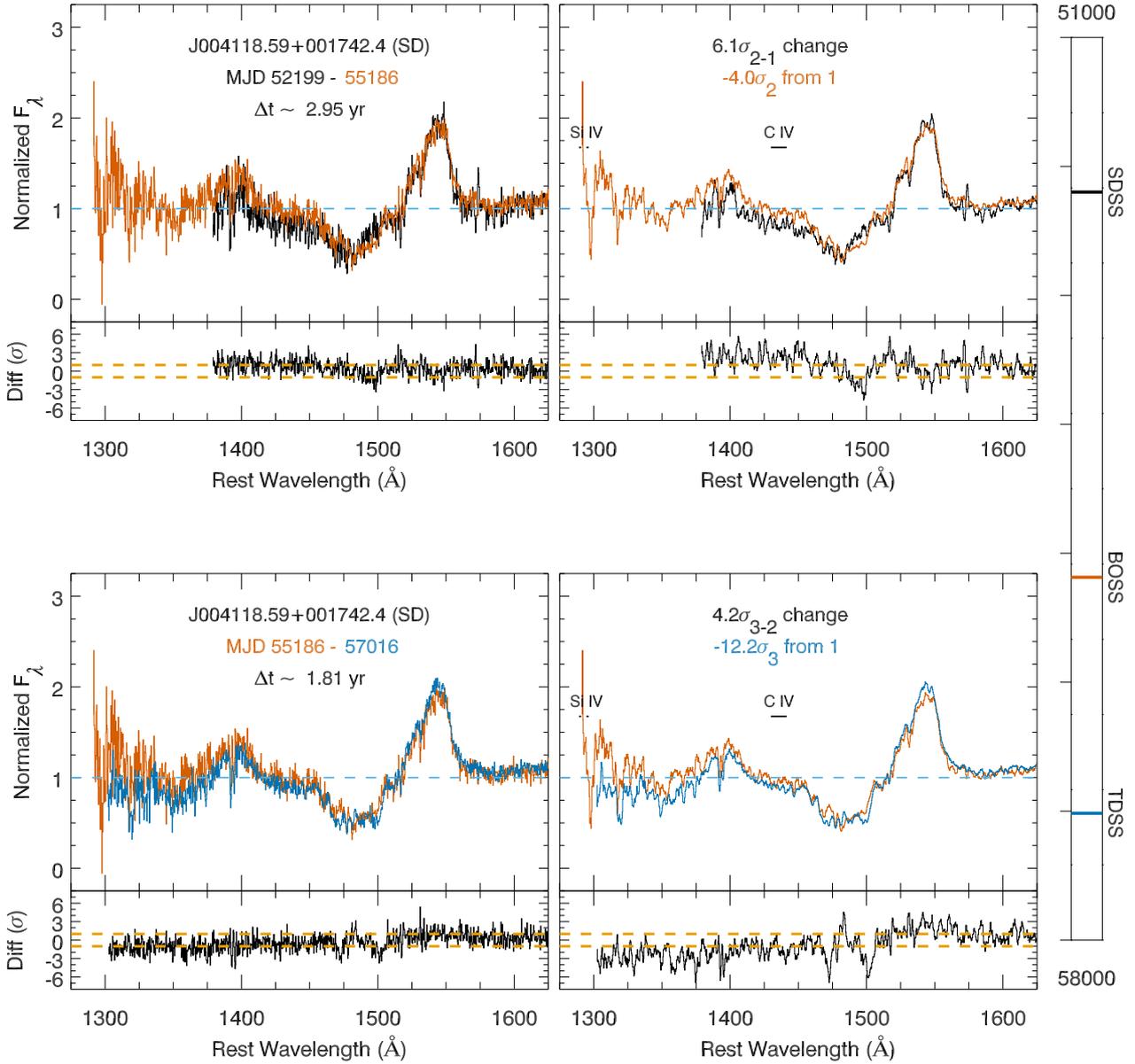}
\caption{Normalized spectrum comparisons showing 24 supplemental cases of disappearing or emerging BALs. Information is displayed in the same format as Fig.~\ref{fig:var}. All the component figures are available in the online, supplementary material (see Fig.~3A).}
\label{fig:sup}
\end{figure*}

The BAL variations that meet conditions A, B, and C listed above are considered part of our supplemental sample of BAL disappearance and emergence events. We also consider events to be supplemental if they satisfy two out of three pristine conditions (i.e., criteria 1 and 2 or criteria 1 and 3) and additionally satisfy the remaining supplemental criterion (either B or C). Fig.~\ref{fig:sup} shows cases of BAL disappearance and emergence from the supplemental sample along with a later observation after the events occurred.\footnote{Fig.~\ref{fig:sup} displays the first component figure from Fig.~3A, which is available in the online, supplementary material (we only refer to Fig.~\ref{fig:sup} hereafter). Fig.~\ref{fig:sup} presents 23 component figures of 23 quasars each with a single, supplemental case of BAL disappearance or emergence, and displays one figure of the quasar J020629.32+004843.0 which shows two supplemental cases of BAL emergence. We also note that two quasars (J023252.80--001351.1 and J111651.98+463508.6) each exhibit a pristine case and supplemental case of BAL disappearance or emergence from the same BAL (shown separately in Fig.~\ref{fig:var} and Fig.~\ref{fig:sup}), and one quasar (J133639.40+514605.3) shows a pristine case and a supplemental case of BAL emergence from different BALs (shown separately in Fig.~\ref{fig:var} and Fig.~\ref{fig:sup}).} 

Upon visual inspection, the comparisons that are considered supplemental exhibit small amplitude variations that are not clearly due to BAL changes (see e.g., in Fig.~\ref{fig:sup}, \mbox{J005215.64+003120.5} and \mbox{J083848.6+411703.9}), show residual absorption that makes it unclear if a BAL disappeared or emerged completely (see e.g., \mbox{J090944.05+363406.7} and \mbox{J093742.27+403351.3),} or exhibit variations that are spurious and do not appear real (see e.g., \mbox{J082059.97+515924.5).} The above examples demonstrate that our supplemental sample consists of marginal cases of BAL disappearance and emergence, and we thus conclude that our pristine sample constitutes a near-complete set of disappearing and emerging BALs in light of the available data. 

The presence of NALs within wavelength intervals where BALs exist can potentially lead to failed detections of pristine cases of BAL disappearance or emergence given the criteria listed above. In order to examine whether NALs significantly contaminate disappearing and emerging BALs in our dataset, we remove criterion 2 and visually inspect all detections of BAL disappearance and emergence by enforcing only criteria 1 and 3. Two BAL quasars \mbox{(J103147.64+575858.0} and \mbox{J120139.34+491327.9)} are contaminated by the presence of NALs in the region where the BAL disappears or emerges. These two cases appear marginal, and we include these objects in our supplemental sample (see Fig.~\ref{fig:sup}). In light of the above evidence, we conclude that NALs do not significantly affect our pristine sample of BAL disappearance and emergence events.

\section{Results}

We apply quantitative criteria to spectra from our 470 BAL quasars and produce a pristine sample of 31 sources with disappearing and/or emerging BALs associated with C\thinspace\textsc{iv} and/or Si\thinspace\textsc{iv} (see Fig.~\ref{fig:var} and Table~\ref{tab:pro}, which lists relevant properties of quasars with pristine cases of disappearing and emerging BALs). We report on the relevant properties of the detected BAL disappearance and emergence cases in Section 4.1 (i.e., frequencies, time-scales, velocities, line strengths, and variability), characteristics of re-emerging and re-disappearing BALs in Sections 4.2 and 4.3, respectively, as well as properties of BAL quasars that host disappearing and emerging BALs in Section 4.4 (i.e., redshifts, absolute \mbox{$i$-band} magnitudes, SMBH masses, bolometric luminosities, Eddington ratios, and radio-loudness parameters). 

\begin{table*}
\centering
\caption{Properties of BAL quasars with pristine disappearing and emerging BALs}
\label{tab:pro}
\begin{tabular}{lccccccccccc}
\hline
BAL quasar & $z$ & $M_i$ & log $L_{\rm{bol}}$ & log $M_{\bullet}$ & log $\frac{L_{\rm{bol}}}{L_{\rm{Edd}}}$ & $R$ & $v_{\rm{BAL}}$ & BI & $\Delta$EW & $\frac{\Delta \rm{EW}}{\langle \rm{EW} \rangle}$ & Ion \\
&& (mag) & (erg s$^{-1}$) & ($M_{\odot}$) &&& (km s$^{-1}$) & (km s$^{-1}$) & (\AA) && \\
\hline
Disappearing BALs \\
\hline
J004022.40+005939.6 $\lozenge$ & 2.56 & --27.1 & 46.5 & 8.7 & --0.3 & 81 & --6900 & 1230 & --10.9 & --2.1 & C\thinspace\textsc{iv} \\
J023252.80--001351.1 $\blacklozenge$ & 2.03 & --27.1 & 46.7 & 9.0 & --0.4 & 0 & --7900 & 50 & --2.8 & --1.6 & Si\thinspace\textsc{iv} \\
J081102.91+500724.2 $\lozenge$ & 1.84  & --26.7 & 46.7 & 9.0 & --0.4 & 224 & --10\,400 & 500 & -6.0 & --1.9 & C\thinspace\textsc{iv} \\
&&&&&&& --9700 & 70 & --2.9 & --1.5 & Si\thinspace\textsc{iv} \\
J091944.53+560243.3 $\blacklozenge$ $\lozenge$ & 1.80 & --26.7 & 46.7 & 9.2 & --0.6 & 0 & --12\,600 & 50 & --3.0 & --2.1 & C\thinspace\textsc{iv} \\
J092522.72+370544.1 & 2.07 & --27.5 & 46.9 & 9.4 & --0.6 & 0 & --14\,800 & 4 & --1.7 & --1.7 & Si\thinspace\textsc{iv} \\
J093535.52+510120.8 & 1.87 & --27.8 & 47.1 & 9.3 & --0.3 & 0 & --35\,100 & 50 & --3.4 & --2.0 & C\thinspace\textsc{iv} \\
J093636.23+562207.4 $\lozenge$ & 1.77 & --27.3 & 46.9 & 9.5 & --0.7 & 0 & --16\,800 & 410 & --8.4 & --2.1 & C\thinspace\textsc{iv} \\
J094456.75+544117.9 $\blacklozenge$ $\lozenge$ & 1.91 & --26.9 & 46.8 & 8.9 & --0.2 & 29 & --14\,300 & 430 & --5.2 & --2.0 & C\thinspace\textsc{iv} \\
J100249.66+393211.0 $\blacklozenge$ & 2.26 & --27.0 & 46.8 & 9.3 & --0.6 & 0 & --6600 & 380 & --5.1 & --1.9 & Si\thinspace\textsc{iv} \\
J111342.27+580407.6 & 2.14 & --27.4 & 46.7 & 10.0 & --1.4 & 0 & --33\,400 & 180 & --4.8 & --2.3 & C\thinspace\textsc{iv} \\
J113754.91+460227.4 $\lozenge$ & 2.11 & --27.0 & 46.6 & 9.0 & --0.5 & 0 & --12\,000 & 560 & --6.8 & --1.9 & C\thinspace\textsc{iv} \\
J125803.13+423812.1 $\lozenge$ & 1.95 & --27.5 & 46.8 & 9.1 & --0.4 & 0 & --19\,600 & 20 & --2.5 & --2.0 & C\thinspace\textsc{iv} \\
J130542.35+462503.4 & 1.83 & --26.5 & 46.6 & 9.2 & --0.7 & 0 & --9600 & 190 & --6.5 & --1.9 & C\thinspace\textsc{iv} \\
J140051.80+463529.9 $\lozenge$ & 1.97 & --28.0 & 47.2 & 9.2 & --0.1 & 0 & --16\,900 & 230 & --4.7 & --2.1 & C\thinspace\textsc{iv} \\
\hline
Emerging BALs \\
\hline
J020629.32+004843.0 $\blacklozenge$ & 2.50 & --27.8 & 47.0 & 10.0 & --1.2 & 0 & --9100 & 280 & 5.2 & 1.9 & Si\thinspace\textsc{iv} \\
J081313.34+414140.6 & 1.70 & --27.4 & 47.1 & 9.6 & --0.6 & 0 & --18\,900 & 1100 & 11.4 & 1.8 & C\thinspace\textsc{iv} \\
&&&&&&& --9600 & 180 & 3.7 & 1.6 & C\thinspace\textsc{iv} \\
J083810.84+430555.7 & 2.12 & --27.6 & 46.8 & 9.4 & --0.7 & 0 & --16\,700 & 830 & 8.7 & 2.0 & Si\thinspace\textsc{iv} \\
J092434.54+423615.1 $\blacklozenge$ & 2.13 & --27.0 & 46.9 & 8.4 &  0.4 & 0 & --9500 & 60 & 2.5 & 1.8 & C\thinspace\textsc{iv} \\
J092507.53+521102.4 & 2.98 & --27.6 & 46.9 & 9.7 & --0.9 & 0 & --21\,100 & 240 & 2.8 & 1.1 & C\thinspace\textsc{iv} \\
J092557.44+410808.8 & 1.80 & --26.4 & 46.5 & 9.5 & --1.1 & 0 & --13\,800 & 200 & 4.5 & 1.7 & C\thinspace\textsc{iv} \\
J093243.25+414230.8 $\blacklozenge$ & 1.98 & --26.7 & 46.5 & 9.2 & --0.8 & 0 & --22\,400 & 190 & 3.0 & 1.5 & C\thinspace\textsc{iv} \\
J095332.92+362552.7 & 2.30 & --27.1 & 46.8 & 8.5 & 0.2 & 0 & --13\,000 & 220 & 4.4 & 2.3 & C\thinspace\textsc{iv} \\
J100318.99+521506.3 & 3.32 & --28.4 & 47.1 & 9.7 & --0.7 & 0 & --14\,600 & 90 & 2.4 & 1.6 & Si\thinspace\textsc{iv} \\
J100906.61+490127.7 & 1.78 & --26.8 & 46.8 & 9.5 & --0.8 & 0 & --15\,800 & 460 & 7.0 & 1.8 & C\thinspace\textsc{iv} \\
J111651.98+463508.6 & 1.89 & --26.6 & 46.6 & 8.6 & --0.1 & 0 & --14\,100 & 40 & 2.8 & 1.9 & C\thinspace\textsc{iv} \\
J111728.75+490216.4 & 2.47 & --27.9 & 47.0 & 8.9 & 0.0 & 0 & --5000 & 170 & 4.0 & 1.9 & Si\thinspace\textsc{iv} \\
J111845.15+504010.4 & 2.34 & --27.6 & 46.8 & -- & -- & 0 & --25\,300 & 120 & 3.7 & 2.2 & C\thinspace\textsc{iv} \\
J121347.74+373726.8 & 1.80 & --27.2 & 46.9 & 9.5 & --0.7 & 0 & --17\,500 & 170 & 3.5 & 2.2 & C\thinspace\textsc{iv} \\
J130542.35+462503.4 & 1.83 & --26.5 & 46.6 & 9.2 & --0.7 & 0 & --16\,400 & 480 & 6.2 & 2.2 & C\thinspace\textsc{iv} \\
J133639.40+514605.3 & 2.22 & --27.0 & 46.6 & 8.8 & --0.3 & 0 & --16\,100 & 40 & 9.9 & 1.8 & C\thinspace\textsc{iv} \\
J134458.82+483457.4 & 2.05 & --26.7 & 46.6 & 9.6 & --1.2 & 0 & --24\,200 & 310 & 3.9 & 1.6 & C\thinspace\textsc{iv} \\
J235859.47--002426.2 & 1.76 & --27.4 & 46.8 & 9.2 & --0.5 & 0 & --17\,200 & 340 & 4.7 & 1.9 & C\thinspace\textsc{iv} \\
\hline
\end{tabular}
\emph{Note.} Columns from left to right include: BAL quasar name, emission-line redshift \citep{hew10}, absolute $i$-band magnitude corrected to $z=2$ \citep{she11}, bolometric luminosity \citep{she11}, SMBH mass \citep{she11}, Eddington ratio \citep{she11}, radio-loudness parameter ($R \equiv f_{\rm{6cm}}/f_{\rm{2500A}}$; \citealt{she11}), disappearing/emerging BAL velocity (see Section 3.2), disappearing/emerging balnicity index (see Section 3.2), disappearing/emerging change in EW (see Section 4.1), disappearing/emerging change in fractional EW (see Section 4.1), and associated ion(s) of the disappearing/emerging BAL(s). Black-filled diamonds indicate quasars with a re-emerging or re-disappearing BAL (see Sections~4.2 and 4.3), and white-filled diamonds indicate sources that undergo a BAL to non-BAL quasar transition (see Section~4.1). Values for $v_{\rm{BAL}}$ and BI for disappearing BALs were measured using the SDSS epoch for each source, while the BOSS epoch was used to measure these quantities for emerging BALs. BAL quasars with $R=0$ were not detected by the FIRST Survey.
\end{table*}

\subsection{Disappearing and emerging BAL properties}

BAL disappearance and emergence are rare events in our BAL quasar sample. A total of 10 out of 31 sources exhibit pristine cases of disappearing C\thinspace\textsc{iv} BALs, 3 out of 31 objects show disappearing Si\thinspace\textsc{iv} BALs, and 1 quasar out of 31 exhibits disappearing C\thinspace\textsc{iv} and Si\thinspace\textsc{iv} BALs at the same velocity (see Table~\ref{tab:pro} and Fig.~\ref{fig:var}). A total of 14 out of 31 pristine objects show C\thinspace\textsc{iv} BAL emergence and 4 out of 31 sources exhibit Si\thinspace\textsc{iv} BAL emergence. There is one BAL quasar where a C\thinspace\textsc{iv} BAL disappears and another C\thinspace\textsc{iv} BAL emerges in the same comparison (see \mbox{J130542.35+462503.4} in Fig.~\ref{fig:var}), and thus the total number of BAL quasars in our pristine sample is 31. The frequencies for BAL quasars in our sample to exhibit C\thinspace\textsc{iv} BAL disappearance and emergence are therefore 2.3$^{+0.9}_{-0.7}$ per cent (11/470) and 3.0$^{+1.0}_{-0.8}$ per cent (14/470), respectively (the 1$\sigma$ error bounds are calculated following \citealt{geh86}). 

Our 11 BAL quasars with pristine cases of disappearing C\thinspace\textsc{iv} BALs and our estimated frequency of 2.3$^{+0.9}_{-0.7}$ per cent are both consistent with \citet{fil12}, who investigated 582 BAL quasars over \mbox{1.1--3.9}~yr and used different criteria and a different sample\footnote{Two of our pristine sources (\mbox{J004022.40+005939.6} and \mbox{J081102.91+500724.2)} are in the pristine sample from \citet{fil12}; nearly all of the remaining quasars from their sample do not meet the criteria to be part of our sample (i.e., having at least 3 epochs separated by $>$~0.1~yr).} to find 11 quasars with pristine cases of disappearing C\thinspace\textsc{iv} troughs (i.e. 1.9$^{+0.8}_{-0.6}$ per cent). Fe\thinspace\textsc{ii} BAL disappearance also appears to be a rare phenomenon; for example, \citet{hal11} detected disappearing Fe\thinspace\textsc{ii} troughs over 0.6--5~yr in J1408+3054, and reported this quasar as the only source to exhibit such dramatic changes out of 24 HiBAL quasars they investigated from the Faint Images of the Radio Sky at Twenty centimeters (FIRST) bright quasar survey \citep{whi00}.  

\begin{figure}
	\centering
	\includegraphics[width=0.45\textwidth]{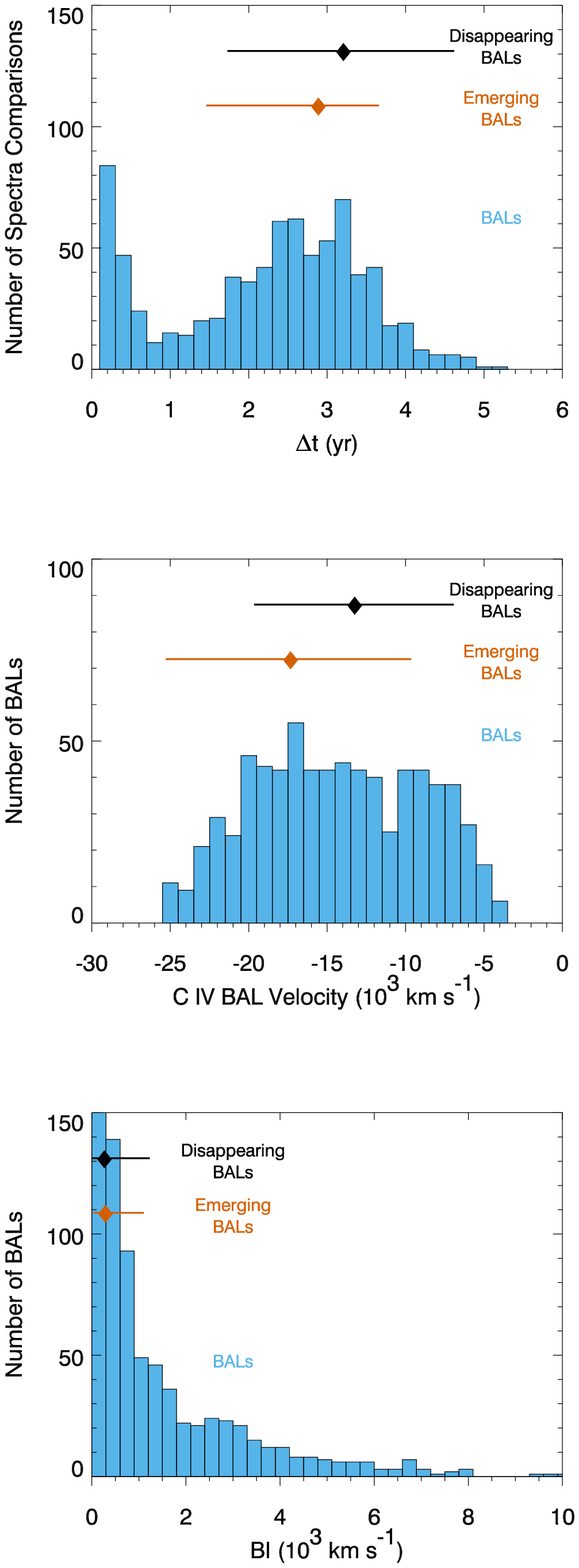}
	\caption{Histograms (blue) for the 470 BAL quasars, displaying measured time-scales between spectra (top panel), C\thinspace\textsc{iv} BAL velocities (middle panel), and C\thinspace\textsc{iv}/Si\thinspace\textsc{iv} balnicity indices (bottom panel). Black and red diamonds above each histogram show locations of average values for disappearing and emerging BALs, respectively. Black and red solid lines extend from minimum to maximum values. Each BAL velocity and balnicity index histogram measurement is an average taken over the available epochs for each BAL quasar.}
	\label{fig:bal}
\end{figure}

BAL disappearance and emergence occur over measured time-scales of \mbox{$\le$ 1.46--4.62}~yr in our pristine sample. Fig.~\ref{fig:bal} presents a histogram of the time-scales we probe using the 470 BAL quasars, and also displays the average and range of times for BALs to disappear and emerge. Spectroscopic comparisons from the 470 BAL quasars probe time-scales between 0.10 and 5.25~yr. To assess the apparent lack of BAL disappearance and emergence over timescales $\la$~1~yr, we conduct two-sample Kolmogorov--Smirnov (K-S) tests comparing the time-scale ($\Delta t$) distributions from the disappearing (D), emerging (E), and full 470 (F) BAL-quasar samples; Table \ref{tab:kst} presents the results. We observe an inconsistency between the D and F distributions (with a K-S probability of 0.02 or 98 per cent confidence) and between the E and F distributions (with 0.03 or 97 per cent confidence). These results indicate that our measured BAL disappearance and emergence time-scales of \mbox{$\le$ 1.46-4.62}~yr are statistically significant within the \mbox{0.10--5.25}~yr range that we probe. Our measured C\thinspace\textsc{iv} BAL disappearance time-scales are broadly consistent with the 1.1--3.9~yr disappearing C\thinspace\textsc{iv} BAL time-scales probed by \citet{fil12} and the \mbox{0.6--5}~yr disappearing Fe\thinspace\textsc{ii} BAL time-scales measured by \citet{hal11}.

 \begin{table*}
 \centering
\caption{Comparisons between BAL quasars with pristine disappearing, emerging, or regular BALs}
\label{tab:kst}
\begin{tabular}{lcccrrrccc}
\hline
Quantity && $N$ &&& Avg. &&& K-S Prob. & \\
& (D) & (E) & (F) & (D) & (E) & (F) & (D,E) & (D,F) & (E,F) \\
\hline
$\Delta t$ (years) & 14 & 18 & 790 & 3.2 $\pm$ 0.2 & 2.9 $\pm$ 0.1 & 2.22 $\pm$ 0.04 & 0.34 & 0.02 & 0.03 \\
$v_{\rm{BAL}}$ (km~s$^{-1}$) & 9 & 15 & 724 & --13\,000 $\pm$ 1000 & --17\,000 $\pm$ 1000 & --14\,400 $\pm$ 200 & 0.09 & 0.59 & 0.06 \\ 
BI (km~s$^{-1}$) & 15 & 19 & 1011 & 300 $\pm$ 100 & 300 $\pm$ 100 & 1180 $\pm$ 50 & 0.57 & 0.02 & 0.006 \\ 
$\Delta$EW (\AA) & 15 & 19 & 706 & --5.0 $\pm$ 0.6 & 5.0 $\pm$ 0.6 & --0.2 $\pm$ 0.1 & 0.70 & $2\times10^{-4}$ & $5\times10^{-7}$ \\
$\Delta$EW/$\langle$EW$\rangle$ & 15 & 19 & 706 & --1.9 $\pm$ 0.1 & 1.8 $\pm$ 0.1 & --0.04 $\pm$ 0.01 & 0.27 & 6$\times10^{-14}$ & 5$\times10^{-17}$ \\
 $z$ & 14 & 18 & 470 & 2.01 $\pm$ 0.06 & 2.2 $\pm$ 0.1 & 2.28 $\pm$ 0.02 & 0.47 & 0.06 & 0.61 \\
 $M_i$ (mag) & 14 & 18 & 470 & --27.0 $\pm$ 0.1 & --26.9 $\pm$ 0.1 & --27.04 $\pm$ 0.03 & 0.91 & 0.21 & 0.52 \\
 log\,$L_{\rm{bol}}$ (erg s$^{-1}$) & 14 & 18 & 470 & 46.83 $\pm$ 0.06 & 46.83 $\pm$ 0.04 & 46.90 $\pm$ 0.01 & 0.62 & 0.59 & 0.88 \\ 
 log\,$M_{\bullet}$ ($M_{\odot}$) & 14 & 17 & 457 & 9.3 $\pm$ 0.1 & 9.5 $\pm$ 0.1 & 9.50 $\pm$ 0.04 & 0.13 & 0.17 & 0.56 \\
 log\,($L_{\rm{bol}}/L_{\rm{Edd}}$) & 14 & 17 & 457 & --0.44 $\pm$ 0.06 & --0.3 $\pm$ 0.2 & --0.32 $\pm$ 0.03 & 0.15 & 0.28 & 0.30 \\
 \hline
 \end{tabular} \\
\emph{Note.} Columns from left to right include: Measured quantity, number of BAL quasars with disappearing (D), emerging (E), and regular (F) BALs used for two-sample K-S tests, average and standard deviation of the mean for the D, E, and F samples, and two-sample, K-S probability between distributions D and E (D,E), D and F (D,F), and E and F (E,F). Measured quantities from top to bottom include: Rest-frame time-scale between pairs of spectra used to detect BAL disappearance/emergence, C\thinspace\textsc{iv} BAL velocity for BALs that lie between $-3000$ and $-27\,600$~km~s$^{-1}$ from the C\thinspace\textsc{iv} BEL, balnicity index, change in BAL EW, change in fractional BAL EW, redshift, absolute $i$-band magnitude, bolometric luminosity, SMBH mass, and Eddington ratio.
\end{table*}

Emerging C\thinspace\textsc{iv} BALs in our pristine sample occur at a marginally higher average radial velocity than disappearing and non-disappearing C\thinspace\textsc{iv} BALs. The C\thinspace\textsc{iv} BALs that emerge between \mbox{$-3000$} and \mbox{$-27\,600$}~km~s$^{-1}$ from the C\thinspace\textsc{iv} BEL do so on average at \mbox{--17\,000}~km~s$^{-1}$, which is 4000~km~s$^{-1}$ higher than the average disappearing C\thinspace\textsc{iv} BAL velocity and 2600~km~s$^{-1}$ higher than the average F sample C\thinspace\textsc{iv} BAL velocity over the same velocity limits (see Table~\ref{tab:kst} and Fig.~\ref{fig:bal}). The E sample C\thinspace\textsc{iv} BAL velocity is inconsistent with the D and F samples at 91 and 94 per cent confidence, respectively (see Table~\ref{tab:kst}). We do not find a significant difference between our D and F BAL velocity distributions, which is inconsistent with \citet{fil12}, who concluded that disappearing C\thinspace\textsc{iv} troughs occur at higher radial velocities than non-disappearing C\thinspace\textsc{iv} troughs in general for their sample. This discrepancy could be related to different velocity limits imposed between studies, and warrants further exploration using a larger sample of disappearing BALs.

Our pristine cases of disappearing and emerging BALs are weak compared to the rest of the BALs in our sample. We quantify the strength of each BAL using the BI formula [see equation (1) in Section 3.2], and the BI value for each pristine disappearing and emerging BAL is listed in Table~\ref{tab:pro}. Disappearing and emerging BALs both have average BI values of \mbox{300$\pm$100}~km~s$^{-1}$, which are on the low end of the BI distribution shown in \mbox{Fig. \ref{fig:bal}}. The average BI value of BALs from the 470 BAL quasars is \mbox{1180$\pm$50}~km~s$^{-1}$. Two-sample K-S tests between the D and F distributions for BI yield an inconsistency at 98 per cent confidence, while the E and F distributions for BI yield an inconsistency with 99.4 per cent confidence (see Table~\ref{tab:kst}). Our result that disappearing BALs are weak compared to BALs from our 470 BAL quasar sample is consistent with previous work (see \citealt{fil12}).

\begin{figure*}
	\centering
	\includegraphics[width=1.90\columnwidth]{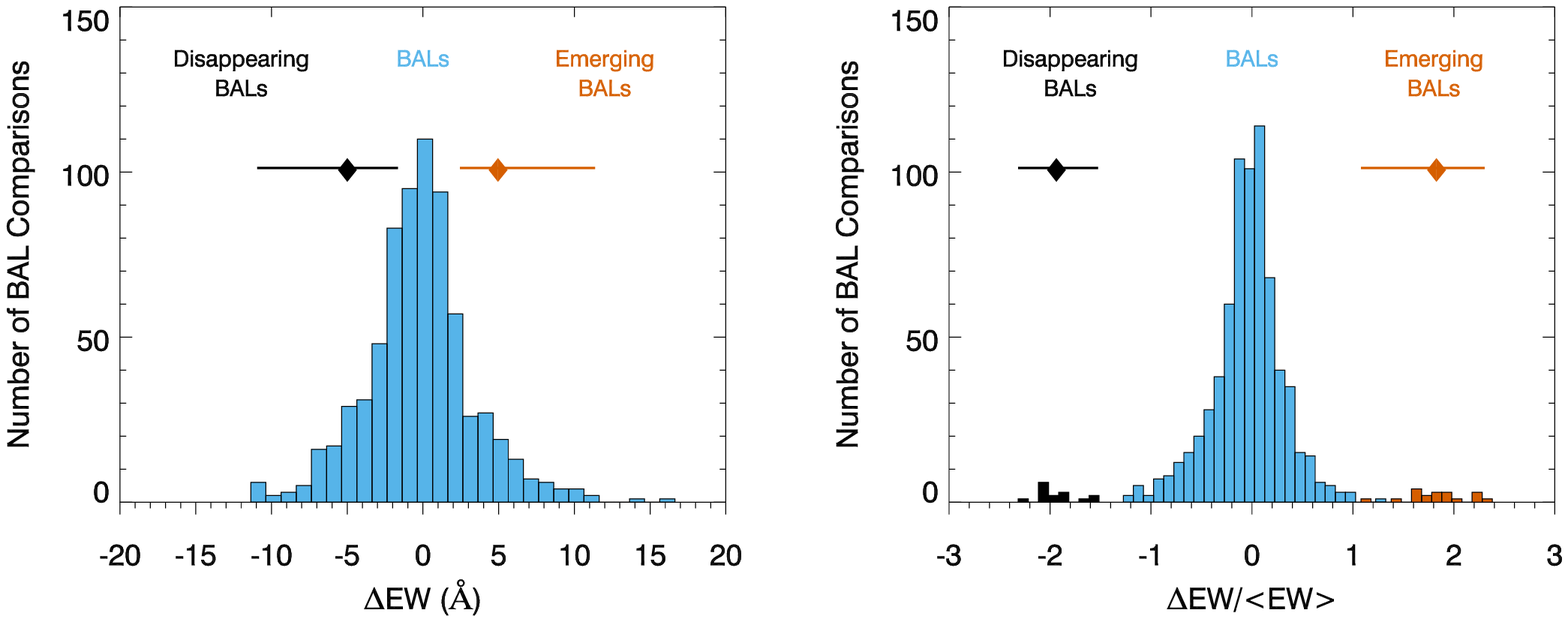}
	\caption{Histograms (blue) for the 470 BAL quasars, displaying change in EW (left panel) and fractional change in EW (right panel) between spectra for C\thinspace\textsc{iv} and Si\thinspace\textsc{iv} BALs. The black and red histograms in the right panel are associated with fractional EW changes for disappearing and emerging BALs, respectively, and are displayed to show the apparent disjointness between the disappearing/emerging and non-disappearing/non-emerging BALs. Other information is displayed in the same format as Fig.~\ref{fig:bal}.}
	\label{fig:bal2}
\end{figure*}

Disappearing and emerging BALs in our pristine sample exhibit fractional EW changes that appear to be distinct from the distribution of fractional EW changes of regular BALs (see Fig.~\ref{fig:bal2}). We calculate the change in EW ($\Delta$EW) and the fractional change in EW ($\Delta$EW/$\langle$EW$\rangle$) of a BAL at times $t_1$ and $t_2$
\begin{equation}
\Delta\textrm{EW}=\textrm{EW}_{2}-\textrm{EW}_{1}
\end{equation}
\begin{equation}
\frac{\Delta\rm{EW}}{\langle\rm{EW}\rangle}=\frac{\rm{EW}_{2}-\rm{EW}_{1}}{0.5\times(\rm{EW}_{2}+\rm{EW}_{1})}
\end{equation}
following \citet{fil13}. The $\Delta$EW values for disappearing and emerging BALs are near the minimum and maximum ends, respectively, of the $\Delta$EW distribution of regular BALs (see Fig.~\ref{fig:bal2} and Table~\ref{tab:pro}). We find highly significant inconsistencies between our D distribution and BALs from our F distribution with negative $\Delta$EW values, and between our E distribution and BALs from our F distribution with positive $\Delta$EW values (both with $>$99.9 per cent confidence). Interestingly, the $\Delta$EW/$\langle$EW$\rangle$ distributions for disappearing and emerging BALs appear to be distinct from the $\Delta$EW/$\langle$EW$\rangle$ distribution of regular BALs (see Fig.~\ref{fig:bal2}); the minimum and maximum $\Delta$EW/$\langle$EW$\rangle$ values for regular BALs are --1.3 and 1.3, respectively, while the average $\Delta$EW/$\langle$EW$\rangle$ values for disappearing and emerging BALs are --1.9 and 1.8, respectively. Two-sample \mbox{K-S} tests yields extremely significant inconsistencies between the D and F distributions and between the E and F distributions for $\Delta$EW/$\langle$EW$\rangle$ (see Table~\ref{tab:kst}). Our results indicate that BAL disappearance/emergence might represent distinct variability events relative to non-disappearing/non-emerging BAL variations. \citet{fil13} concluded that BAL disappearance events appear to be extremes of BAL variability and not a distinct phenomenon (see their fig. 14), however there were only a small number (3) of cases available for their analysis. Future studies with larger samples will be able to further investigate the apparent distinction between disappearing/emerging and non-disappearing/non-emerging BALs.

Some emerging and disappearing C\thinspace\textsc{iv} BALs in our sample are correlated with other C\thinspace\textsc{iv} BALs in the same set of spectra. A total of 8 out of 13 emerging C\thinspace\textsc{iv} BALs are associated with increases from a non-emerging C\thinspace\textsc{iv} BAL at a different velocity. This behavior is only marginally significant however, as the binomial distribution probability (see e.g., \citealt{bev03}) for 8 out of 13 `successes' for uncorrelated variability is 16 per cent. There are three disappearing C\thinspace\textsc{iv} BALs with a non-disappearing C\thinspace\textsc{iv} BAL at a different velocity, and in one source the non-disappearing BAL decreases in amplitude during the disappearance event (see J111342.27+580407.6 in Fig. \ref{fig:var}). \citet{fil12} found that in 11 out of 12 cases (0.3 per cent probability) there were non-disappearing C\thinspace\textsc{iv} BALs that weakened with disappearing C\thinspace\textsc{iv} troughs in their sample. We cannot make a proper comparison with this result due to our small number (3) of relevant comparison sources.

We find no emerging C\thinspace\textsc{iv} BALs in our sample with associated emerging Si\thinspace\textsc{iv} absorption, but observe marginal, correlated behavior between disappearing Si\thinspace\textsc{iv} BALs and associated non-disappearing C\thinspace\textsc{iv} BALs. Out of seven comparisons with wavelength coverage of an emerging C\thinspace\textsc{iv} BAL and the interval where Si\thinspace\textsc{iv} absorption would emerge at the same velocity, no comparisons show emerging Si\thinspace\textsc{iv} absorption with the emerging C\thinspace\textsc{iv} BAL. All four disappearing Si\thinspace\textsc{iv} BALs have associated C\thinspace\textsc{iv} BALs that decrease in amplitude during the Si\thinspace\textsc{iv} BAL disappearance event (see \mbox{J023252.80--001351.1}, \mbox{J081102.91+500724.2}, \mbox{J092522.72+370544.1}, and \mbox{J100249.66+393211.0} in Fig.~\ref{fig:var}); this behavior yields a binomial probability of 6 per cent.  

Disappearing C\thinspace\textsc{iv} BALs in our sample usually produce a BAL to non-BAL quasar transition during the disappearance. For the 11 quasars with a disappearing C\thinspace\textsc{iv} BAL, 8 sources do not show any BALs after the C\thinspace\textsc{iv} BAL disappears while 3 objects exhibit non-disappearing C\thinspace\textsc{iv} broad absorption at a distinct velocity. BAL to non-BAL transitioning quasars are marked with white-filled diamonds next to their names in Fig.~\ref{fig:var} and Table~\ref{tab:pro}. The frequency of BAL to non-BAL quasar transition given our dataset is thus 1.7$^{+0.8}_{-0.6}$ per cent (8/470), which is consistent with \citet{fil12} who estimated a transition frequency of 1.7$^{+0.7}_{-0.5}$ per cent with their 582 BAL quasars.

\subsection{Re-emerging BAL properties}

\begin{figure*}
	\centering
	\includegraphics[width=2.00\columnwidth]{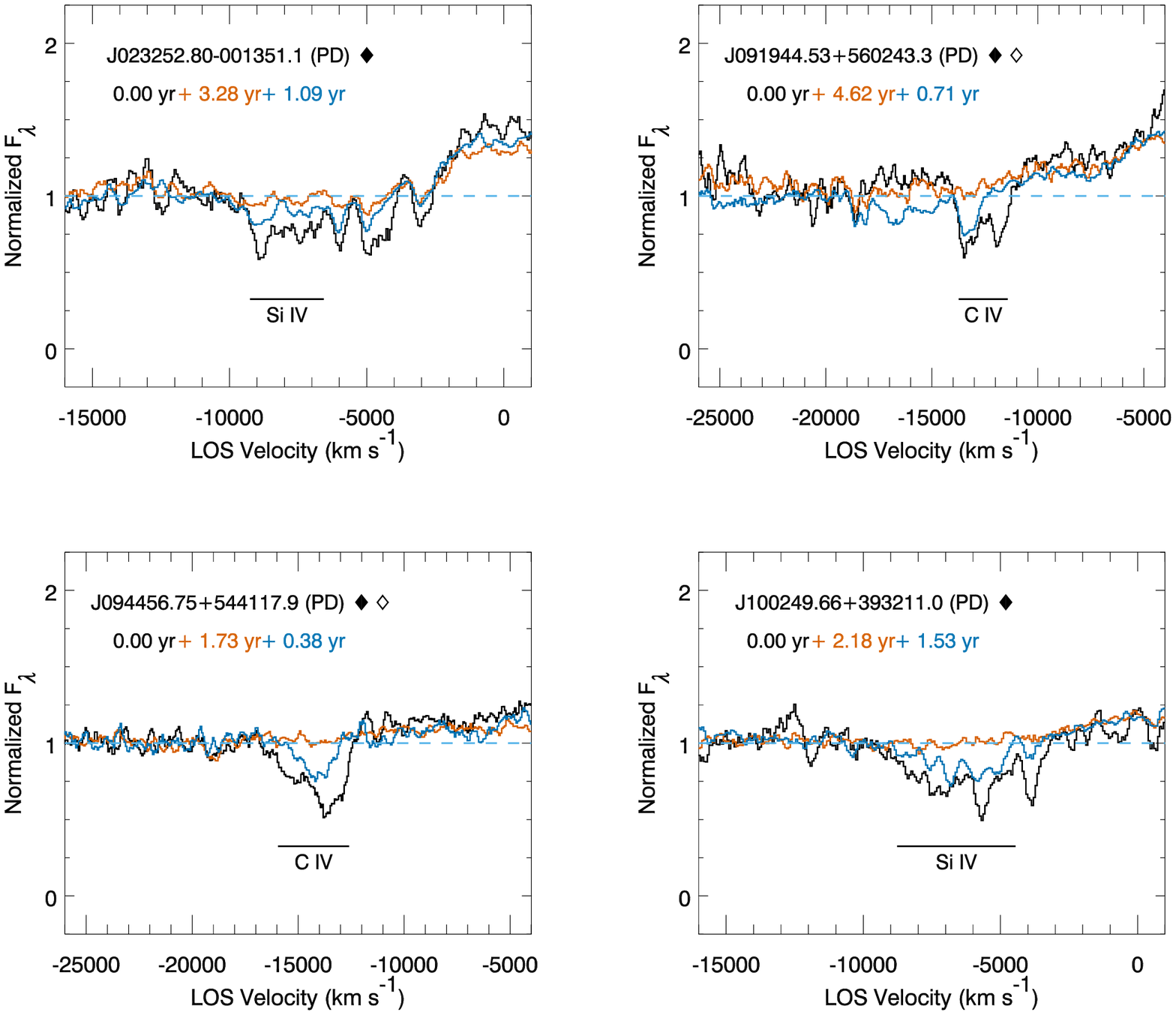}
	\caption{Four pristine cases of disappearing (PD) BALs that re-emerge in a later observation. Each panel lists the BAL quasar name followed by three time durations that correspond to the three plotted spectra: the first epoch (black) at 0.00 yr, the second epoch (red) at the number of years since the first spectrum, and the third epoch (blue) at the number of years since the second spectrum. Spectra are smoothed using a 5-pixel boxcar for better presentation. The associated ion is shown below each re-emerging BAL, and the solid line above each ion extends over the range of the BAL that re-emerged.}
	\label{fig:pdr}
\end{figure*}

We detect 14 BAL quasars with pristine BAL disappearance events that occur over measured time-scales \mbox{$\le$ 1.73--4.62}~yr; four of these objects show a BAL re-emergence event\footnote{We define a BAL re-emergence event to be one where the average flux $F_3$ of the re-emerging trough is less than the continuum minus 4$\sigma_3$ (see criterion 2 in Section 3.2) and the average flux difference between the two spectra is $>7\sigma_{2-3}$ (see criterion 3 in Section 3.2).} in the later observation. The four re-emerging BALs are shown in Table~\ref{tab:pro} and Fig.~\ref{fig:var} with black-filled diamonds next to their names, and are also displayed by themselves in Fig. \ref{fig:pdr} over a more limited velocity range for detailed visual inspection. Two of the four re-emerging BALs are associated with C\thinspace\textsc{iv}, while the other two are associated with Si\thinspace\textsc{iv} (see Fig. \ref{fig:pdr}). The disappearing BALs in the remaining 10 sources do not re-emerge in the available later epoch. 

The four re-emerging troughs exhibit a smaller flux difference during their re-emergence than during their initial BAL disappearance; this behavior might be caused by measured time-scale differences between SDSS/BOSS comparisons and BOSS/TDSS comparisons in our dataset (see Fig.~\ref{fig:var} and \ref{fig:pdr}). For the two quasars with re-emerging C\thinspace\textsc{iv} BALs we observe a BAL to non-BAL quasar transition during the initial C\thinspace\textsc{iv} BAL disappearance; the C\thinspace\textsc{iv} BAL re-emergence events do not formally constitute non-BAL to BAL quasar transitions since the re-emerging troughs are formally mini-BALs (see Section 3.2 and Fig. \ref{fig:pdr}). The two re-emerging Si\thinspace\textsc{iv} BALs are correlated with C\thinspace\textsc{iv} BALs at the same velocity; the C\thinspace\textsc{iv} BALs decrease in strength during the Si\thinspace\textsc{iv} BAL disappearance and increase in strength during the Si\thinspace\textsc{iv} BAL re-emergence (see Fig.~\ref{fig:var}). The re-emerging Si\thinspace\textsc{iv} BAL in \mbox{J100249.66+393211.0} is the only re-emerging trough that is formally considered a BAL (see Section 3.2 and Fig.~\ref{fig:pdr}). All four troughs shown in Fig.~\ref{fig:pdr} re-emerge at roughly the same velocity and have notable similarities in their kinematic profiles as the BALs before the disappearance event (see Appendix A for more details on individual objects).

\subsection{Re-disappearing BAL properties}

\begin{figure}
	\centering
	\includegraphics[width=1.00\columnwidth]{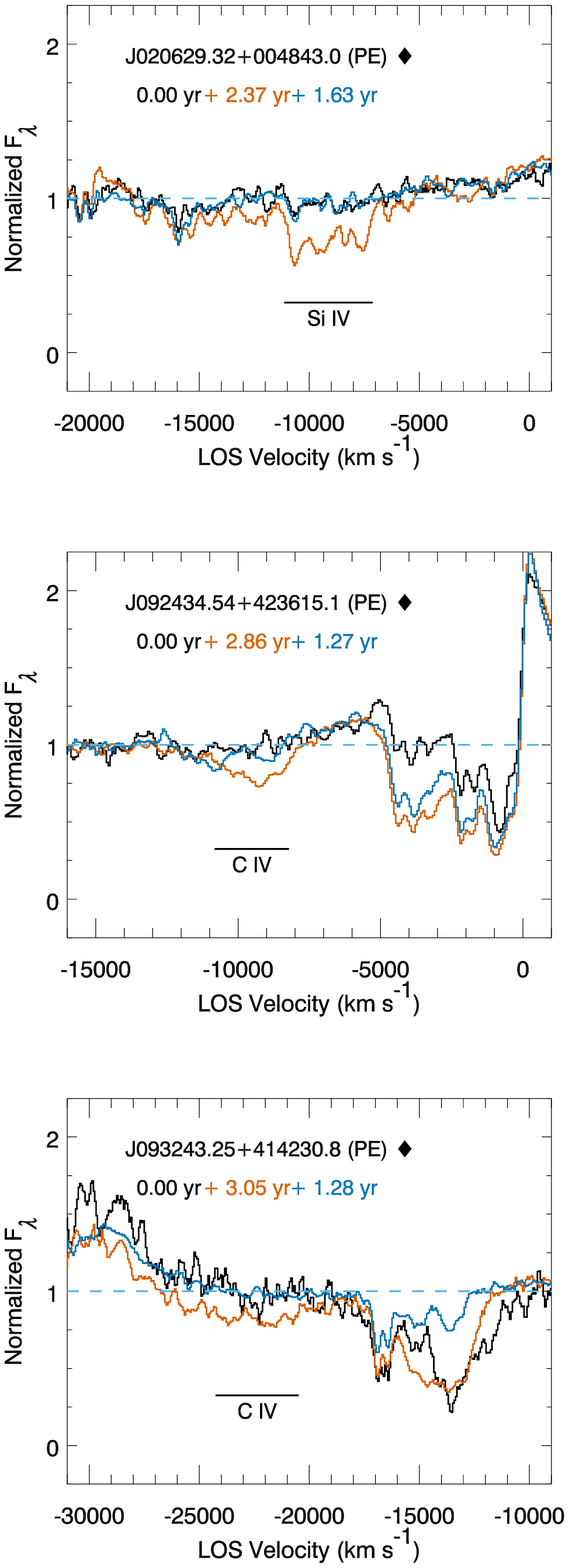}
	\caption{Three pristine cases of emerging (PE) BALs that re-disappear in the available later observation. Information is displayed in the same format as Fig.~\ref{fig:pdr}.}
	\label{fig:per}
\end{figure}

We find 18 BAL quasars with pristine BAL emergence events that occur over measured time-scales \mbox{$\le$ 1.46--3.66}~yr, and three of these sources exhibit a BAL re-disappearance event\footnote{We define a BAL re-disappearance event to be one where the average flux $F_3$ of the re-disappearing BAL is greater than the continuum minus 7$\sigma_3$ (see criterion B in Section 3.2) and the average flux difference between the two spectra is $>7\sigma_{3-2}$ (see criterion 3 in Section 3.2). We adopt the 1--7$\sigma_3$ threshold instead of 1--4$\sigma_3$ in order to examine more completely re-disappearance cases in our sample.} in the final epoch. The three re-disappearing BAL quasars are shown in Table~\ref{tab:pro} and Fig.~\ref{fig:var} with black-filled diamonds next to their names, and are also presented individually in Fig. \ref{fig:per} over a more limited velocity range for detailed visual inspection. Two of the three re-disappearing BALs are associated with C\thinspace\textsc{iv}, and the remaining one is associated with Si\thinspace\textsc{iv} (see Fig.~\ref{fig:per}). Six of the 15 remaining emerging BALs decrease in strength in the available later observation, while the other 9 emerging BALs do not change their profiles in a significant way in the later epoch. It is noteworthy that no emerging BALs in our sample continue to strengthen in the later observation, and this behavior warrants further investigation using a larger sample.

The two re-disappearing C\thinspace\textsc{iv} BALs are correlated with C\thinspace\textsc{iv} BALs at different velocities; in both cases the C\thinspace\textsc{iv} BALs at different velocities decrease in strength during the C\thinspace\textsc{iv} BAL re-disappearance (see Fig. \ref{fig:per}). The re-disappearing Si\thinspace\textsc{iv} BAL in J020629.32+004843.0 is correlated not only with a C\thinspace\textsc{iv} BAL at the same velocity, but also with another C\thinspace\textsc{iv} BAL at a different velocity; the C\thinspace\textsc{iv} BALs increase in strength during the Si\thinspace\textsc{iv} BAL emergence and decrease in strength during the Si\thinspace\textsc{iv} BAL re-disappearance (see Fig. \ref{fig:var}). See Appendix A for more details on individual sources. \citet{fil12} reported an example of a C\thinspace\textsc{iv} BAL re-disappearance event associated with the quasar J093620.52+004649.2 (see their fig.~5); this source does not meet our criteria established in Section 3.2 due to the possible residual absorption before the BAL emergence event occurs.

In Section 5 we use the properties of (re-)disappearing and (re-)emerging BALs to discuss the origin of the variability in these absorbers, and consider the implications for BAL-outflow lifetimes, sizes, and distances from the SMBH.  

\subsection{Properties of BAL quasars with disappearing and emerging BALs}

\begin{figure}
	\centering
	\includegraphics[width=0.47\textwidth]{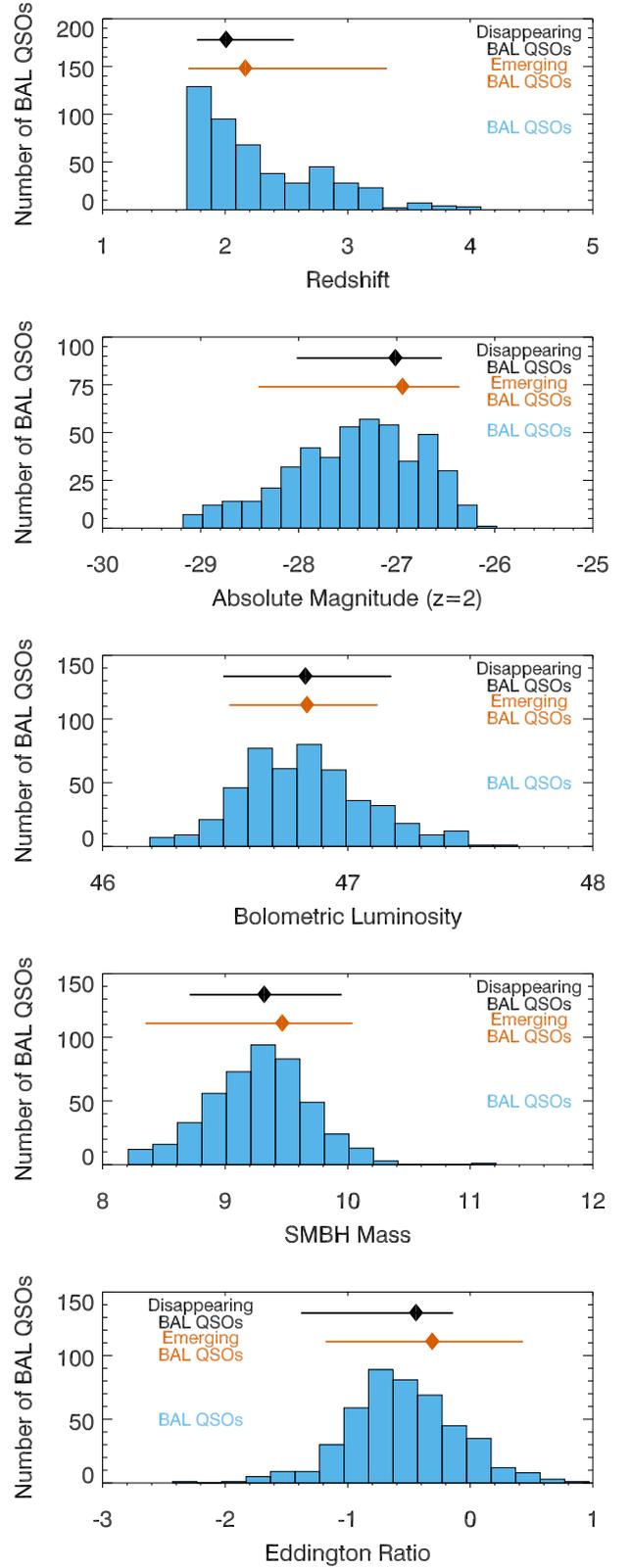}
	\caption{Histograms (blue) displaying redshifts (top panel), absolute $i$-band magnitudes, bolometric luminosities, SMBH masses, and Eddington ratios (bottom panel) for the 470 BAL-quasar sample. Information is displayed in the same format as Fig.~\ref{fig:bal}.}
	\label{fig:qso}
\end{figure}

Fig.~\ref{fig:qso} displays histograms for the 470 BAL quasars including redshift $z$, absolute $i$-band magnitude $M_i$, bolometric luminosity log\,($L_{\rm{bol}}$), SMBH mass log\,($M_{\bullet}$), and Eddington ratio log\,($L_{\rm{bol}}/L_{\rm{Edd}}$), and it also indicates representative values for our BAL quasars with pristine cases of disappearing and emerging BALs. Table~\ref{tab:pro} lists these properties for each BAL quasar with pristine cases of disappearing and emerging BALs, and Table~\ref{tab:kst} presents two-sample K-S test results between the pristine disappearing (D), emerging (E), and full 470 (F) BAL-quasar distributions to compare the quasar properties between quasars hosting disappearing/emerging and non-disappearing/non-emerging BALs.

The average redshift of BAL quasars with disappearing BALs is 2.01, which is 0.27 lower than the average redshift for the 470 BAL quasars (see Fig.~\ref{fig:qso}). We conduct two-sample K-S tests between $M_i$, log\,($L_{\rm{bol}}$), log\,($M_{\bullet}$), and log\,($L_{\rm{bol}}/L_{\rm{Edd}}$) distributions from our D, E, and F BAL quasar samples, and the largest inconsistency is with 87 per cent confidence (see Table~\ref{tab:kst}). Absolute magnitudes and bolometric luminosities presented in Fig. \ref{fig:qso} and Table~\ref{tab:pro} were derived using representative quasar spectral energy distributions (SEDs) (see \citealt{ric06}; \citealt{she11}), and single-epoch, virial SMBH mass estimates have uncertainties of up to 0.5~dex (see \citealt{she13}). In light of the above evidence and the number of K-S tests performed, we conclude that there are no statistically significant differences between the disappearing, emerging, and 470 BAL quasar properties within our sample. Our results are consistent with previous work \citep{fil12,fil13}.  

There are three radio-loud\footnote{We consider BAL quasars with radio-loudness parameters \mbox{$R>10$} to be radio-loud (see e.g., \citealt{jia07}). Other studies sometimes describe quasars with $R=10-100$ to be `radio-intermediate.'} BAL quasars with disappearing BALs and no radio-loud BAL quasars with emerging BALs in our pristine sample (see Table 1). Out of the 467 sources in our sample that were observed by the FIRST survey, 35 objects are radio-loud and 411 sources are radio-quiet (i.e. $R<10$). The BAL quasars that are radio-loud have $R$ values that range from $R=11$ to $R=598$ and have $R=78$ on average. Our detections of disappearing BALs associated with both radio-loud and radio-quiet BAL quasars is consistent with results from \citet{fil12} (see their section 4.2). It is interesting that we find no radio-loud BAL quasars with emerging BALs, and a larger sample will be able to assess this possibility statistically.

Our 470 BAL-quasar sample contains 71 LoBAL quasars that are defined by the presence of Al\thinspace\textsc{iii} broad absorption (see footnote 5) in addition to C\thinspace\textsc{iv} and Si\thinspace\textsc{iv} BALs. There is one quasar that is categorized as a LoBAL out of the 31 sources in our pristine sample (see J130542.35+462503.4 in Fig.~\ref{fig:var}). This LoBAL quasar is the only object in our pristine sample that exhibits a BAL disappearance and emergence event in the same comparison. The Al\thinspace\textsc{iii} BAL in J130542.35+462503.4 disappears at the same velocity as the disappearing C\thinspace\textsc{iv} BAL and disappearing Si\thinspace\textsc{iv} absorption (see Fig.~\ref{fig:var}). The C\thinspace\textsc{iv} BAL emergence event in this quasar appears to be accompanied by emerging Si\thinspace\textsc{iv} absorption at the same velocity (see Fig.~\ref{fig:var}). The small number of LoBAL quasars in our pristine sample is consistent with the result that disappearing and emerging BALs are generally weak compared to BALs in general (see Section 4.1 and \citealt{fil12}). The presence of Al\thinspace\textsc{iii} broad absorption at the same velocity as C\thinspace\textsc{iv} has been shown to be indicative of large C\thinspace\textsc{iv} BAL equivalent widths (see e.g., \citealt{fil14}). 

\section{Discussion}

\subsection{The origin of disappearing/emerging BALs}

There are two plausible scenarios for BAL disappearance and emergence over our measured multi-year time-scales. The first considers outflowing gas responding to fluctuations in the ionizing radiation field, which causes a change in optical depth and thus a change in the BAL profile. The second scenario involves outflowing gas moving transversely to our LOS, which produces a change in coverage of the background light source and hence a change in the BAL profile. Regardless of origin, in Section 5.1.1 we discuss the constraints that disappearing BALs place on BAL lifetimes along the LOS. In Sections 5.1.2 and 5.1.3 we discuss a number of situations that yield evidence supporting ionization-change and transverse-motion scenarios, respectively, and provide examples that are consistent with these situations using BAL quasars from our pristine sample. \mbox{Table~\ref{tab:int}} lists the pristine cases of BAL disappearance and emergence in our sample, and categorizes the evidence we find that supports one of the two scenarios driving BAL variability.

\subsubsection{BAL lifetimes}

Detecting BALs that disappear over a measured time interval allows us to obtain a rough estimate for the lifetime $t_{\rm{BAL}}$ over which BALs are visible along our LOS. This lifetime could be the time required for a BAL structure to cross our LOS or the time between ionization events significant enough to reduce the EW of C\thinspace\textsc{iv} below our detection limits (in the latter case, the BAL gas remains along our LOS but the BAL is no longer visible in C\thinspace\textsc{iv}). In either case, we constrain $t_{\rm{BAL}}$ by rearranging the equation \mbox{$f=\langle \Delta t \rangle/t_{\rm{BAL}}$}, where $f$ is the fraction of BALs that disappear over the time interval $\langle \Delta t \rangle$.\footnote{Since disappearing BALs are found to be weak compared to non-disappearing BALs (see Section 4.1 and \citealt{fil12}), we caution that this method of estimating $t_{\rm{BAL}}$ might not be applicable to all BALs.} This calculation is directly following the calculation in \citet{fil12} (see their section 4.1). The number of detected pristine cases of BAL disappearance out of our 470 BAL-quasar sample allows for an estimate of $f$, and our measured time-scales for BAL disappearance place an upper limit for $\langle \Delta t \rangle$.

\begin{table*}
\centering
\caption{Interpretations of pristine disappearing/emerging BALs}
\label{tab:int}
\begin{tabular}{ccccll}
\hline
BAL quasar & BAL  & Scenario & Category &  Primary Evidence & Secondary Evidence \\
\hline
Disappearing BALs \\
\hline
J004022.40+005939.6 & C\thinspace\textsc{iv} & TM & Strong & Saturated C\thinspace\textsc{iv} BAL \\
J023252.80--001351.1 & Si\thinspace\textsc{iv} & IC & Very Strong & Re-emerging, preserved Si\thinspace\textsc{iv} BAL & Variable C\thinspace\textsc{iv} BEL \\
&&&&& CSS continuum variation \\
J081102.91+500724.2 & C\thinspace\textsc{iv} & TM & Strong & Saturated C\thinspace\textsc{iv} BAL \\
& Si\thinspace\textsc{iv} &&& \\
J091944.53+560243.3 & C\thinspace\textsc{iv} & ? && \\
J092522.72+370544.1 & Si\thinspace\textsc{iv} & IC & Weak & CSS continuum variation \\
J093535.52+510120.8 & C\thinspace\textsc{iv} & ? && \\
J093636.23+562207.4 & C\thinspace\textsc{iv} & IC & Strong & Wide, disappearing C\thinspace\textsc{iv} BAL \\
J094456.75+544117.9 & C\thinspace\textsc{iv} & IC & Strong & Re-emerging C\thinspace\textsc{iv} BAL \\
J100249.66+393211.0 & Si\thinspace\textsc{iv} & IC & Very Strong & Re-emerging, preserved Si\thinspace\textsc{iv}/C\thinspace\textsc{iv} BAL \\
J111342.27+580407.6 & C\thinspace\textsc{iv} & IC & Weak & CSS continuum variation \\
J113754.91+460227.4 & C\thinspace\textsc{iv} & TM & Strong & Saturated C\thinspace\textsc{iv} BAL \\
J125803.13+423812.1 & C\thinspace\textsc{iv} & IC & Weak & CSS continuum variation \\
J130542.35+462503.4 & C\thinspace\textsc{iv} & TM & Strong & Saturated C\thinspace\textsc{iv} BAL \\
J140051.80+463529.9 & C\thinspace\textsc{iv} & IC & Weak & Variable C\thinspace\textsc{iv} BEL \\
\hline
Emerging BALs \\
\hline
J020629.32+004843.0 & Si\thinspace\textsc{iv} & IC & Very Strong & Preserved C\thinspace\textsc{iv} BAL \\
J081313.34+414140.6 & C\thinspace\textsc{iv} & IC & Strong & Wide, emerging C\thinspace\textsc{iv} BALs \\
& C\thinspace\textsc{iv} &&& \\
J083810.84+430555.7 & Si\thinspace\textsc{iv} & IC & Weak & Variable C\thinspace\textsc{iv} BEL \\
J092434.54+423615.1 & C\thinspace\textsc{iv} & IC & Weak & CSS continuum variation \\
J092507.53+521102.4 & C\thinspace\textsc{iv} & ? && \\
J092557.44+410808.8 & C\thinspace\textsc{iv} & IC & Weak & CSS continuum variation \\
J093243.25+414230.8 & C\thinspace\textsc{iv} & IC & Weak & Variable C\thinspace\textsc{iv} BEL \\
J095332.92+362552.7 & C\thinspace\textsc{iv} & IC & Weak & Variable C\thinspace\textsc{iv} BEL \\
J100318.99+521506.3 & Si\thinspace\textsc{iv} & ? && \\
J100906.61+490127.7 & C\thinspace\textsc{iv} & ? && \\
J111651.98+463508.6 & C\thinspace\textsc{iv} & ? && \\
J111728.75+490216.4 & Si\thinspace\textsc{iv} & ? && \\
J111845.15+504010.4 & C\thinspace\textsc{iv} & ? && \\
J121347.74+373726.8 & C\thinspace\textsc{iv} & IC & Weak & Variable C\thinspace\textsc{iv} BEL & CSS continuum variation \\
J130542.35+462503.4 & C\thinspace\textsc{iv} & TM & Strong & Saturated C\thinspace\textsc{iv} BAL \\
J133639.40+514605.3 & C\thinspace\textsc{iv} & ? && \\
J134458.82+483457.4 & C\thinspace\textsc{iv} & IC & Strong & Coordinated, variable C\thinspace\textsc{iv} BALs \\
J235859.47--002426.2 & C\thinspace\textsc{iv} & ? && \\
\hline
\end{tabular} \\
\emph{Note.} Columns from left to right include: BAL quasar name, disappearing/emerging BAL indicated by the associated ion (see Figs.~\ref{fig:var},~\ref{fig:pdr}, and~\ref{fig:per}), scenario that is most consistent with each BAL disappearance/emergence event [i.e., transverse motions (TM), ionization changes (IC), or indeterminate (?)], strength of evidence supporting each listed scenario (i.e., Very Strong, Strong, or Weak), and specific  primary and secondary evidence supporting each listed scenario (see Sections~5.1.2 and~5.1.3 for detailed discussion). The quasars J081102.91+500724.2 and J081313.34+414140.6 have two BALs that collectively provide evidence for one scenario.
\end{table*}

Using the equation above, we obtain an upper limit for $t_{\rm{BAL}}$ by utilizing the six disappearing C\thinspace\textsc{iv} BALs that do not re-emerge in the later observation and are also associated with BAL to non-BAL quasar transitions (sources in Table~\ref{tab:pro} with only white-filled diamonds next to their names). The average measured time for C\thinspace\textsc{iv} BAL disappearance in these six cases is $\langle \Delta t \rangle \le 3.28$~yr (see Fig.~\ref{fig:var}), and the fraction of cases out of our 470 BAL-quasar sample is $f=1.3^{+0.8}_{-0.5}$ per cent. These two measurements yield a limit of $t_{\rm{BAL}} \la 250^{+160}_{-100}$~yr for the BAL lifetime along the LOS. We can obtain another upper limit for $t_{\rm{BAL}}$ if we consider BALs that disappear then re-emerge in the later observation. We detect two C\thinspace\textsc{iv} BALs that disappear then re-emerge over an average time-scale of $\langle \Delta t \rangle \le 3.72$~yr (see Fig.~\ref{fig:var}), and the fraction of these cases out of 470 BAL quasars is thus  $f=0.4^{+0.6}_{-0.3}$ per cent. These two measurements therefore yield a limit for the BAL lifetime along our LOS to be $t_{\rm{BAL}}\la900^{+1200}_{-600}$~yr. 

Our BAL-lifetime estimates are somewhat higher than the estimate from \citet{fil12}, who used 21 cases of C\thinspace\textsc{iv} BAL disappearance and a similar calculation to infer a BAL lifetime along the LOS of 109$^{+31}_{-22}$~yr. Conversely, \citet{fil13} utilized the equivalent widths of non-disappearing C\thinspace\textsc{iv} and Si\thinspace\textsc{iv} BALs, which are found to be generally stronger than disappearing BALs, and inferred a BAL lifetime along the LOS of 1600$^{+2100}_{-900}$~yr for their sample of 291 BAL quasars. Our nominal upper limits in combination with previous work imply BAL lifetimes along the LOS to be on the order of $\la$100--1000~yr using a sample of BAL quasars with disappearing and re-emerging BALs (see also \citealt{gib08}; \citealt{hal11}). We note that our BAL lifetime limits are strictly only applicable for relatively weak BALs, as we find that disappearing/emerging BALs in our sample exhibit generally low BI values compared to the BI distribution from our full 470 BAL-quasar sample (see bottom panel of Fig.~\ref{fig:bal}). The LOS lifetime for an individual BAL can exist well outside this range, however, as we detect three cases of BAL re-disappearance over multi-year time-scales (see Fig.~\ref{fig:per}).

\subsubsection{Evidence for ionization changes}

Observing BALs with notable similarities in their kinematic profiles before and after re-emergence events provides very strong evidence supporting ionization-change scenarios if the outflows are non-axisymmetric; transverse motions of gas that is not uniform cannot easily produce similar line profiles before and after BALs re-emerge (see BALs in Table~\ref{tab:int} with ``Very Strong" evidence in support of the ionization-change scenario). The BAL quasar J100249.66+393211.0 exhibits a C\thinspace\textsc{iv} BAL which possesses a similar line profile and depth before the Si\thinspace\textsc{iv} BAL disappearance at the same velocity (MJD 53033) and after the re-emergence (MJD 57461; see Fig.~\ref{fig:var}). The C\thinspace\textsc{iv} BAL at the same velocity as the re-disappearing Si\thinspace\textsc{iv} BAL in \mbox{J020629.32+004843.0} recovers it original profile and depth after the re-disappearance event (see Fig.~\ref{fig:var}). The fact that C\thinspace\textsc{iv} broad absorption remained present throughout the Si\thinspace\textsc{iv} re-emergence/re-disappearance events in the above two sources is very strong evidence against transverse-motion scenarios for simple outflow models. J023252.80--001351.1 and J100249.66+393211.0 each show a re-emerging Si\thinspace\textsc{iv} trough that kinematically resembles the Si\thinspace\textsc{iv} BAL profile before the disappearance event (see Fig.~\ref{fig:var} and \ref{fig:pdr} and Appendix~A). J094456.75+544117.9 exhibits a re-emerging C\thinspace\textsc{iv} trough that kinematically resembles the C\thinspace\textsc{iv} BAL that initially disappeared at the same velocity (see Fig.~\ref{fig:var} and \ref{fig:pdr} and Appendix A); since this re-emerging trough does not appear to preserve its kinematic profile as well as the ``Very Strong" candidates listed in Table~\ref{tab:int}, we categorize this behaviour as ``Strong" evidence in support of the ionization-change model.

Coordinated BAL variability over a large velocity interval (e.g., greater than $\sim$10\,000~km~s$^{-1}$) is more easily explained by situations involving an ionization change; such behavior in the transverse-motion scenario would require coordinated motions of gas that would presumably exist at a large range of distances from the SMBH (see e.g., \citealt{ham11}; \citealt{cap13}; \citealt{gri15}; \citealt{wan15}). Fig. \ref{fig:var} displays a C\thinspace\textsc{iv} BAL emerging over a $\sim$14\,500 km s$^{-1}$ range in J081313.37+414140.6, and a C\thinspace\textsc{iv} BAL disappearing over a $\sim$11\,600 km s$^{-1}$ interval in J093636.23+562207.4. The BAL quasar J134458.82+483457.4 shows an emerging C\thinspace\textsc{iv} BAL and other C\thinspace\textsc{iv} BALs that increase in line depth over a $\sim$15\,500 km s$^{-1}$ range (see Fig.~\ref{fig:var}). We consider these three cases as ``Strong" evidence in support of the ionization-change scenario (see Table~\ref{tab:int}).

Significant BEL variability that originates from the photoionized broad line region is an indicator of fluctuations in the ionizing radiation field that could produce ionization changes within BAL outflows.\footnote{Time delays and other effects make it difficult to make direct comparisons between BEL and BAL variations, and we are thus only discussing BEL variability as evidence for fluctuations in the ionizing radiation field. We therefore consider the following BALs with associated BEL changes as ``Weak" evidence in support of the ionization-change model (see Table~\ref{tab:int}).} We detect a significant decrease in the C\thinspace\textsc{iv} BEL strength during the Si\thinspace\textsc{iv} BAL-disappearance event in J023252.80--001351.1 as well as during the Si\thinspace\textsc{iv} BAL-emergence event in J083810.84+430555.7 (see Fig.~\ref{fig:var}). Some quasars in our sample exhibit disappearing and emerging C\thinspace\textsc{iv} BALs that do not have detectable Si\thinspace\textsc{iv} absorption at the same velocity, indicating the possibility that in these cases C\thinspace\textsc{iv} might be more responsive to ionization changes because the BAL gas is too highly ionized (see \citealt{wan15}) or because the gas is optically thin in C\thinspace\textsc{iv} (see \citealt{fil14} as well as discussion in Section 5.1.3 in reference to \citealt{ham98}). Fig.~\ref{fig:var} presents a disappearing or emerging C\thinspace\textsc{iv} BAL with no associated Si\thinspace\textsc{iv} absorption and with significant C\thinspace\textsc{iv} BEL changes in J093243.25+414230.8, J095332.92+362552.7, J121347.74+373726.8, and J140051.80+463529.9.

\begin{figure}
	\centering
	\includegraphics[width=1.00\columnwidth]{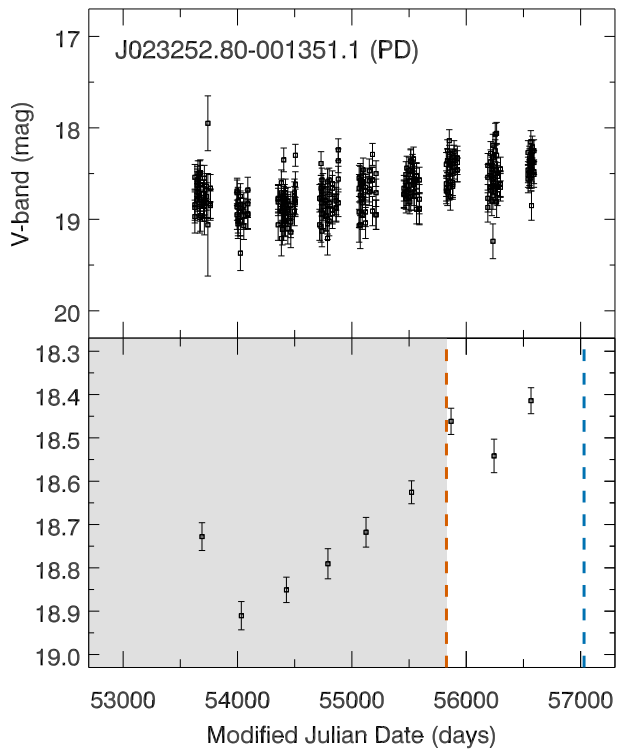}
		\caption{Synthesized CSS $V$-band light curves showing notable photometric variations for BAL quasars with pristine cases of disappearing (PD) and emerging (PE) BALs. Each pair of connected panels corresponds to the same quasar. The top panel in each pair lists the source name and displays the synthesized \mbox{$V$-band} measurements. Each bottom panel displays mean $V$-band measurements from the top panel in 1-year bins, along with standard deviations of the mean for each bin. Vertical black, red, and blue dashed lines lie at the same MJD and possess the same colour as spectra from the corresponding BAL quasar displayed in Fig.~\ref{fig:var}. Grey regions correspond to time intervals where BAL disappearance/emergence is detected. All the component figures are available in the online, supplementary material (see Fig.~9A).}
		\label{fig:css}
\end{figure}

Detecting continuum variations within our available wavelength coverage might be indicative of fluctuations in ionizing radiation at shorter wavelengths if the two wavelength regimes exhibit correlated variability. Fig.~\ref{fig:css} presents\footnote{Fig.~\ref{fig:css} displays the first of eight component figure from Fig.~9A, which can be viewed in the online, supplementary material (we refer to Fig.~\ref{fig:css} hereafter).} notable\footnote{The CSS light curves for BAL quasars from our pristine sample not presented in Fig.~\ref{fig:css} exhibit small magnitude changes (i.e. $\la$0.1~mag) or show too large uncertainties for detecting a trend. We also note that the synthesized $V$-band fluctuations in Fig.~\ref{fig:css} do not appear to be unique compared to BAL quasars more generally; we selected 31 random objects from our 470 BAL-quasar sample, and found 6 sources with CSS light curves that show similar, significant trends with similar variation amplitudes.} CSS, synthesized $V$-band light curves for BAL quasars with pristine cases of disappearing and emerging BALs. The $V$-band fluctuations in Fig.~\ref{fig:css} exhibit amplitudes between 0.2--0.7~mag over time intervals where BAL disappearance or emergence occurs (grey regions). The amplitude of these synthesized $V$-band changes cannot be entirely explained by variations from BAL disappearance/emergence, BAL variability, and BEL changes; the average flux differences over \mbox{1300--1600}~\AA \ in these sources produce changes $\la0.1$~mag for all objects in Fig.~\ref{fig:css} except for J121347.74+373726.8.\footnote{The BAL and BEL variations in this quasar produce a 0.2~mag \emph{weakening} in flux, but CSS detects a $\sim$0.2~mag \emph{strengthening} in flux over the same time period.}  Inspection of normalized spectroscopic comparisons over 1600--2200~\AA \ for these sources reveals no significant variations of the C\thinspace\textsc{iii}] $\lambda$1909 BEL, which is the only prominent feature within this interval for these objects. The detected $V$-band variations in Fig.~\ref{fig:css} are therefore likely to arise at least partly from fluctuations in continuum radiation. 

It is noteworthy that, for each of the BAL quasars shown in Fig. \ref{fig:css}, there is either a significant brightening or dimming in flux during the measured time that we detect either a disappearing or emerging BAL. We detect a $\sim$0.5 and 0.3~mag flux brightening during the BAL disappearance events in \mbox{J023252.80--001351.1} and \mbox{J092522.72+370544.1,} respectively, and observe a $\sim$0.7 and 0.3~mag flux dimming during the BAL disappearance events in J111342.27+580407.6 and J125803.13+423812.1, respectively (see Fig.~\ref{fig:css} in relation to Fig.~\ref{fig:var}). We observe similar brightening and dimming trends (i.e., between $\sim$0.2--0.4~mag) for BAL quasars with emerging BALs in Fig.~\ref{fig:css} (see also Fig.~\ref{fig:var}). These flux changes are significant given that the standard deviations of the \mbox{1-year}, average $V$-band measurements are $\la0.05$~mag. Our results are consistent with \citet{wan15} who detected 50 cases of correlated behavior between absorption-line emergence or disappearance and continuum changes, providing further support for the ionization-change scenario. Due to the uncertain connection between the observed continuum variations and fluctuations in the ionizing radiation field, we consider the above cases as ``Weak" evidence supporting the ionization-change interpretation (see Table~\ref{tab:int}).

\subsubsection{Evidence for transverse motions}

Detection of disappearing or emerging BALs with saturated components is strong evidence supporting scenarios where outflows traverse our LOS; fluctuations in the ionizing radiation field cannot easily induce changes to absorbers with sufficiently high optical depths (see BALs in Table~\ref{tab:int} with ``Strong" evidence in support of the transverse-motion scenario). \citet{ham98} examined modeling using the code \textsc{cloudy} \citep{fer96} to determine theoretical line optical depths for constant-density, photoionized clouds that have solar abundances and square absorption profiles with velocity widths of \mbox{10\,000}~km~s$^{-1}$. Fig. 6 from \citet{ham98} shows that the predicted C\thinspace\textsc{iv} BAL optical depths are at least a factor of 10 higher than the Si\thinspace\textsc{iv} BAL optical depths for ionization parameters between \mbox{$-1 < \textrm{log} \ U < +1$.}

We detect four disappearing C\thinspace\textsc{iv} BALs that have simultaneously disappearing Si\thinspace\textsc{iv} absorption at the same velocity, indicating possible saturation in the C\thinspace\textsc{iv} BALs (see, in Fig.~\ref{fig:var}, \mbox{J004022.40+005939.6,} \mbox{J081102.91+500724.2,} \mbox{J113754.91+460227.4,} and J130542.35+462503.4). J130542.35+462503.4 also exhibits an emerging C\thinspace\textsc{iv} BAL with Si\thinspace\textsc{iv} absorption at the same velocity (see Fig.~\ref{fig:var}). It is noteworthy that each C\thinspace\textsc{iv} BAL mentioned above has a comparable\footnote{The C\thinspace\textsc{iv} BAL depths are $\sim$0.1 larger than their corresponding Si\thinspace\textsc{iv} BAL depths for the four quasars mentioned above.  This difference in line depth is much smaller than other quasars in our sample with C\thinspace\textsc{iv} and Si\thinspace\textsc{iv} BALs at the same velocity, where the difference is $\sim$0.5 (see e.g., in Fig.~\ref{fig:var}, J02062932+004843.0, J023252.80--001351.1, and J100249.66+393211.0).} line depth to its corresponding Si\thinspace\textsc{iv} absorption line, providing evidence of saturation in these absorbers and therefore the transverse-motion scenario. Large C\thinspace\textsc{iv} BAL optical depths would be `hidden' within the non-zero line depths we measure if the absorbers only partially cover their background light sources (see e.g., \citealt{ham99}; \citealt{ara99,ara01}).

In order to predict broadly the degree of saturation in the disappearing C\thinspace\textsc{iv} BALs mentioned above, we measure an average, normalized flux of \mbox{$F\sim0.75$} for the corresponding Si\thinspace\textsc{iv} BALs before they disappear (see Fig.~\ref{fig:var}). Assuming complete coverage $C$ of the background light source (i.e., $C=1$) yields a Si\thinspace\textsc{iv} BAL optical depth of $\tau_{\rm{Si IV}}=0.3$ (using $F=e^{-\tau}$), which yields a predicted C\thinspace\textsc{iv} BAL optical depth of at least $\tau_{\rm{CIV}}=3.0$ using the theoretical modeling from \citet{ham98}. However the comparable line depths observed between the Si\thinspace\textsc{iv} and C\thinspace\textsc{iv} BALs is indicative of partial coverage of the background light source (i.e., $C<1$), which would lead to higher predicted C\thinspace\textsc{iv} BAL optical depths. If the absorbers, for example, have a coverage of $C=0.4$, then the estimated Si\thinspace\textsc{iv} BAL optical depth would become $\tau_{\rm{SiIV}}=1.0$ [using $F=Ce^{-\tau}+(1-C)$] and the predicted C\thinspace\textsc{iv} BAL optical depth would become at least $\tau_{\rm{CIV}}=10.0$; C\thinspace\textsc{iv} absorption with this optical depth might not easily respond to changes in the ionizing radiation field to produce the highly significant, BAL-disappearance events that we observe. 

In light of the discussion in this section and in Section 5.1.2, we find evidence across our sample of disappearing/emerging BALs that collectively supports both the ionization-change and transverse-motion scenarios. Inspection of Table~\ref{tab:int} reveals that seven disappearing/emerging BALs provide ``Very Strong" or ``Strong" evidence supporting ionization-change models, while five disappearing/emerging BALs provide ``Strong" evidence in support of the transverse-motion interpretation. These results are consistent with the finding in section 4.7 of \citet{fil13} that at least $56\pm7$ per cent of \ion{C}{IV} BAL troughs exhibit variations consistent with being caused by ionization changes. We do not find evidence of transverse motions and ionization changes occurring in the same BAL quasar in our dataset. However, since our constraints for a number of trough variations are weak, we cannot rule out that such `mixed' behaviour is present. Larger sample sizes will be needed to assess these possibilities statistically. Below we consider each of the two models and discuss implications for the physical properties of BAL outflows in our pristine sample. 

\subsection{BAL outflows undergoing ionization changes}

We utilize our disappearing and emerging BAL sample\footnote{We do not include measurements from the BAL quasars \mbox{J004022.40+005939.6,} \mbox{J081102.91+500724.2,} \mbox{J113754.91+460227.4,} and \mbox{J130542.35+462503.4} in the subsequent calculations, as these sources exhibit BAL disappearance/emergence that provides strong evidence for transverse-motion scenarios in our sample (see Section 5.1.3 and Table~\ref{tab:int}).} to examine outflows responding to ionizing-radiation fluctuations in order to obtain a nominal estimate for the outflow distance $r$ from the SMBH \\
\begin{equation}
 r=\sqrt{\frac{L_{\rm{ion}}}{4\pi U n_{\rm{e}} c}} 
 \end{equation} \\
 where $L_{\rm{ion}}$ is the ionizing photon luminosity above 13.6~eV, $U$ is the ionization parameter, $n_{\rm{e}}$ is the electron density, and $c$ is the speed of light. We also use our re-emerging and re-disappearing BALs, which in this model would be dominated by changes in optical depth, to obtain a nominal estimate of the radial outflow size $D_{\rm{rad}}$ relative to our LOS \\
\begin{equation}
D_{\rm{rad}}=\bigg(\frac{m_{\rm{e}} c}{\pi e^{2} f \lambda}\bigg) \frac{\tau_{\rm{BAL}} \Delta v_{\rm{BAL}}}{n_{\rm{e}}}
\end{equation} \\
where $m_{\rm{e}}$ is the electron mass, $e$ is the fundamental unit of charge, $f$ is the oscillator strength, $\lambda$ is the rest wavelength, $\tau_{\rm{BAL}}$ is the BAL optical depth, and $\Delta v_{\rm{BAL}}$ is the velocity width of the BAL. Equation (8) is derived by rearranging the relation between column density $N$ and optical depth [i.e., $N \equiv \int n_{\rm{e}} ds = \sigma \int \tau(v) dv$, where $\sigma$ is the absorption cross section], and assumes constant-density absorbers and an average BAL optical depth (see section 1 from \citealt{sav91}). Table~\ref{tab:mea} presents relevant measurements for these and subsequent calculations from our pristine cases of BAL disappearance/emergence.  

\begin{table*}
\centering
\caption{Measurements of pristine disappearing/emerging BALs}
\label{tab:mea}
\begin{tabular}{lccccc}
\hline
BAL quasar & $\Delta t$ & $\Delta v_{\rm{BAL}}$ & $F$ & $\Delta A$ & Ion \\
& (yr) & (km s$^{-1}$) &&& \\
\hline
Disappearing BALs \\
\hline
J004022.40+005939.6 & 2.24 & 6000 & 0.66 & 0.35 & C\thinspace\textsc{iv} \\
J023252.80--001351.1 $\blacklozenge$ & 3.28 & 2400 & 0.75 & 0.22 & Si\thinspace\textsc{iv} \\
J081102.91+500724.2 & 3.50 & 4100 & 0.70 & 0.29 & C\thinspace\textsc{iv} \\
&& 2200 & 0.69 & 0.32 & Si\thinspace\textsc{iv} \\
J091944.53+560243.3 $\blacklozenge$ & 4.62 & 2300 & 0.75 & 0.31 & C\thinspace\textsc{iv} \\
J092522.72+370544.1 & 2.61 & 2100 & 0.83 & 0.13 & Si\thinspace\textsc{iv} \\
J093535.52+510120.8 & 3.92 & 2400 & 0.73 & 0.38 & C\thinspace\textsc{iv} \\
J093636.23+562207.4 & 4.61 & 3400 & 0.79 & 0.21 & C\thinspace\textsc{iv} \\
J094456.75+544117.9 $\blacklozenge$ & 1.73 & 3400 & 0.70 & 0.32 & C\thinspace\textsc{iv} \\
J100249.66+393211.0 $\blacklozenge$ & 2.18 & 3900 & 0.74 & 0.27 & Si\thinspace\textsc{iv} \\
J111342.27+580407.6 & 3.74 & 3000 & 0.71 & 0.34 & C\thinspace\textsc{iv} \\
J113754.91+460227.4 & 2.94 & 4300 & 0.69 & 0.30 & C\thinspace\textsc{iv} \\
J125803.13+423812.1 & 3.02 & 2300 & 0.79 & 0.18 & C\thinspace\textsc{iv} \\
J130542.35+462503.4 & 3.17 & 3500 & 0.78 & 0.36 & C\thinspace\textsc{iv} \\
J140051.80+463529.9 & 3.36 & 3400 & 0.74 & 0.26 & C\thinspace\textsc{iv} \\
\hline
Emerging BALs \\
\hline 
J020629.32+004843.0 $\blacklozenge$ & 2.37 & 3600 & 0.72 & 0.24 & Si\thinspace\textsc{iv} \\
J081313.34+414140.6 & 3.35 & 11\,200 & 0.79 & 0.19 & C\thinspace\textsc{iv} \\
&& 2900 & 0.72 & 0.24 & C\thinspace\textsc{iv} \\ 
J083810.84+430555.7 & 3.24 & 6200 & 0.74 & 0.30 & Si\thinspace\textsc{iv} \\
J092434.54+423615.1 $\blacklozenge$ & 2.86 & 2600 & 0.81 & 0.18 & C\thinspace\textsc{iv} \\
J092507.53+521102.4 & 3.00 & 3300 & 0.76 & 0.20 & C\thinspace\textsc{iv} \\
J092557.44+410808.8 & 3.21 & 3200 & 0.70 & 0.28 & C\thinspace\textsc{iv} \\
J093243.25+414230.8 $\blacklozenge$ & 3.05 & 3800 & 0.82 & 0.17 & C\thinspace\textsc{iv} \\
J095332.92+362552.7 & 2.15 & 4400 & 0.82 & 0.18 & C\thinspace\textsc{iv} \\
J100318.99+521506.3 & 2.54 & 3100 & 0.83 & 0.18 & Si\thinspace\textsc{iv} \\
J100906.61+490127.7 & 3.60 & 3700 & 0.61 & 0.49 & C\thinspace\textsc{iv} \\
J111651.98+463508.6 & 1.46 & 3200 & 0.83 & 0.16 & C\thinspace\textsc{iv} \\
J111728.75+490216.4 & 3.12 & 2600 & 0.70 & 0.28 & Si\thinspace\textsc{iv} \\
J111845.15+504010.4 & 3.31 & 3500 & 0.80 & 0.24 & C\thinspace\textsc{iv} \\
J121347.74+373726.8 & 2.10 & 3300 & 0.80 & 0.20 & C\thinspace\textsc{iv} \\
J130542.35+462503.4 & 3.17 & 4800 &  0.76 & 0.29 & C\thinspace\textsc{iv} \\
J133639.40+514605.3 & 2.53 & 5300 & 0.61 & 0.36 & C\thinspace\textsc{iv} \\
J134458.82+483457.4 & 3.29 & 3600 & 0.76 & 0.19 & C\thinspace\textsc{iv} \\
J235859.47--002426.2 & 3.66 & 4600 & 0.80 & 0.19 & C\thinspace\textsc{iv} \\
 \hline
\end{tabular}\\
\emph{Note.} Columns from left to right include: BAL quasar name, measured BAL disappearance/emergence time-scale, BAL velocity width, BAL flux before disappearance or after emergence, change in normalized BAL flux during disappearance/emergence, and associated ion. Black-filled diamonds correspond to re-emerging or re-disappearing BALs.
\end{table*}

We calculate $L_{\rm{ion}}$ using a segmented power law from \citet{ham11} to serve as a typical BAL-quasar SED (see their appendix A for more details), and adopting an average luminosity at 1350~\AA, in erg~s$^{-1}$, of log\,$L_{1350}=46.2$ to serve as the normalization (measurements taken from \citealt{she11}). The estimate for $L_{\rm{ion}}$, in photons s$^{-1}$, is log\,$L_{\rm{ion}}=56.4$. We adopt an ionization parameter of log\,$U=0$ based on results from \citet{wan15}, who found correlated variations between C\thinspace\textsc{iv}, Si\thinspace\textsc{iv}, and N\thinspace\textsc{v} BALs and the continuum in 452 quasars and conducted photoionization modeling to constrain log $U \ga 0$ for their sample.

We determine a lower limit for $n_{\rm{e}}$ by assuming that the absorbers exist sufficiently far from the SMBH for the recombination time to be much shorter than the ionization time of the gas, and that the gas is not too highly ionized relative to C\thinspace\textsc{iv} and Si\thinspace\textsc{iv} [i.e., $n_{\rm{e}}\ga 1/(\alpha_{\rm{rec}} \Delta t$), where $\alpha_{\rm{rec}}$ is the recombination-rate coefficient and $\Delta t$ is an upper limit for the BAL-variability timescale (see Table~\ref{tab:mea}); see \citealt{cap13} and \citealt{gri15} for details]. We calculate $\alpha_{\rm{rec}}$ using the \mbox{\textsc{chianti}} atomic database version 8.0 (\citealt{der97}; \citealt{del15}) with a nominal BAL-outflow temperature of 20\,000~K; the coefficients for C\thinspace\textsc{iv} and Si\thinspace\textsc{iv} are $1.5\times10^{-11}$ and $1.7\times10^{-11}$~cm$^{3}$~s$^{-1}$, respectively, and we adopt the average value for our calculation. The recombination rate used in combination with the average measured timescale of $\Delta t\le3.04$~yr for pristine BAL disappearance and emergence thus yields an electron-density lower limit of $n_{\rm{e}} \ga 650$~cm$^{-3}$. The above values allow for a nominal upper limit for the outflow distance from the SMBH to be on the order of $r\la100$~pc, assuming the conditions outlined above. Using extreme measurements from our sample yields minimum and maximum nominal upper limits for $r$ to be on the order of $\la$ 100~pc and 1~kpc, respectively. 

BAL line-depth variations are dictated by changes in optical depth under the ionization change scenario; measured line depths of re-emerging and re-disappearing BALs therefore likely trace nearly the entire radial extent of the flow where C\thinspace\textsc{iv} and Si\thinspace\textsc{iv} arises, since the entire BAL profile in these cases disappears or emerges presumably from fluctuations in the ionizing radiation field. For our calculation of $D_{\rm{rad}}$ using equation (8), we use average $f$ and $\lambda$ values for C\thinspace\textsc{iv} and Si\thinspace\textsc{iv} using measurements from \citet{mor03}. BAL optical depths $\tau_{\rm{BAL}}$ are determined using average, normalized fluxes $F$ across the BAL (see Table~\ref{tab:mea}), and assume complete coverage of the continuum source (i.e., $\tau_{\rm{BAL}}=-\rm{ln}$\,$F$).\footnote{Assuming complete coverage (i.e., $C=1$) results in a lower limit on $D_{\rm{rad}}$, since absorbers that partially cover the continuum source (i.e., $C<1$) would result in higher BAL optical depths and thus larger radial outflow sizes (see also footnote 23).} BAL widths extend from the minimum to maximum velocity of the BAL, and we measure an average value of $\Delta v_{\rm{BAL}}=3100$~km~s$^{-1}$ for our re-emerging and re-disappearing BALs (see sources with black-filled diamonds in Table~\ref{tab:mea}). The average BAL flux is $F=0.76$ for the re-emerging and re-disappearing BALs before they disappear or after they emerge, respectively (see objects with black-filled diamonds in Table~\ref{tab:mea}). The above considerations yield a nominal lower limit for the radial outflow size for BALs in our sample to be on the order of $D_{\rm{rad}}\ga1\times10^{-7}$~pc.\footnote{BAL outflows with $D_{\rm{rad}}=1\times10^{-7}$~pc and with radial velocities extending over $\Delta v_{\rm{BAL}}=3100$~km~s$^{-1}$ may indicate that the absorbers are distributed in small clumps over a larger radial extent than our $D_{\rm{rad}}$ lower limit. It is also plausible that these outflows exhibit larger radial sizes if the gas is highly ionized relative to C\thinspace\textsc{iv} and Si\thinspace\textsc{iv}.} Limiting measurements from our sample yield minimum and maximum nominal lower limits for the radial outflow size to be on the order of $\ga1\times10^{-8}$ and $1\times10^{-6}$~pc, respectively.

We have utilized a large BAL quasar sample to place order-of-magnitude limits on the location and radial extent of disappearing/emerging BAL outflows; previous work has mainly focused on constraining the properties of non-disappearing/non-emerging BAL winds. Our range of nominal BAL-outflow distance upper limits of \mbox{$r\la100-1000$}~pc are consistent with models where BAL gas accelerated off the accretion disc travels out to kpc-scale distances from the SMBH and interacts with the surrounding ISM (e.g., \citealt{fau12a}; \citealt{fau12b}). Our BAL-outflow distance limits, obtained using disappearing/emerging BALs, are consistent with previous non-disappearing/non-emerging BAL-variability studies using HiBAL and LoBAL quasars that have attributed BAL variations to ionization changes and have constrained the flows to exist within a few hundred parsecs of the SMBH (\citealt{bar94}; \citealt{viv12}; \citealt{gri15}). Our estimates are also consistent with previous studies using photoionization models and density-sensitive absorption lines to estimate BAL-outflow locations. \citet{ara13} utilized the O\thinspace\textsc{vi} absorption doublet and \textsc{cloudy} models to estimate a BAL-outflow distance of 3~kpc, while \citet{bor13} and \citet{cha15} both used the S\thinspace\textsc{iv} doublet to estimate an outflow distance of 100~pc and between 100--2000~pc from the SMBH, respectively.

Our nominal radial BAL-outflow size lower limit of $D_{\rm{rad}}\ga1\times10^{-7}$~pc (0.02~au) is consistent with \citet{ham13}, who placed a nominal upper limit of \mbox{$\la 1\times10^{-4}$}~pc for the radial-outflow size in their sample of 8 mini-BAL quasars (see also discussion at the end of Section 5.3). We emphasize that our radial BAL-outflow size constraint utilizes pristine cases of BAL disappearance/emergence in our sample, whereas the size limits from \citet{ham13} made use of non-disappearing/non-emerging mini-BAL variations in combination with \textsc{cloudy} photoionization models. The above estimates support models where outflows are dense and geometrically small compared to their distance from the SMBH; \citet{ham13} derived a maximum radial filling factor of $\la 5\times10^{-5}$ allowed by their data, which is consistent with our estimates. It is therefore plausible given this discussion that the weak, disappearing/emerging BAL outflows that we probe consist of small substructures. It is unclear as to whether strong BALs consist of similar, compact flow geometries, and this possibility warrants future investigation.

\subsection{BAL outflows traversing our LOS}

We utilize measurements from our disappearing and emerging BAL sample\footnote{We do not include the BAL quasars \mbox{J020629.32+004843.0,} \mbox{J023252.80--001351.1,} \mbox{J094456.75+544117.9,} and \mbox{J100249.66+393211.0} in the subsequent calculations, as these objects exhibit BAL re-emergence/re-disappearance that provides the strongest evidence for ionization-change scenarios in our sample (see Section 5.1.2 and Table~\ref{tab:int}).} and consider outflows moving transversely to our LOS at Keplerian\footnote{Moravec et al. (submitted) reports that the Keplerian speed is a reasonable first guess to the transverse outflow speed, since models predict that BAL outflows are launched from a rotating accretion disc.} speeds in order to obtain a nominal estimate for the outflow distance $r$ from the SMBH \\
\begin{equation}
 r=\frac{GM_{\bullet}}{v_{\rm{trans}}^{2}}
 \end{equation} \\
 where $G$ represents the gravitational constant, $M_{\bullet}$ is the SMBH mass, and $v_{\rm{trans}}$ is the transverse outflow speed. We also use the re-disappearing BALs from J092434.54+423615.1 and J093243.25+414230.8, which are presumed to be dominated by changes in coverage under this scenario, in order to nominally estimate the transverse outflow size $D_{\rm{trans}}$ relative to our LOS \\
\begin{equation}
D_{\rm{trans}}=v_{\rm{trans}}\Delta t'
\end{equation} \\
where $\Delta t'$ is the upper limit for the timescale of a \emph{complete} BAL re-disappearance event (i.e., the emergence time plus the re-disappearance time). 

We estimate $v_{\rm{trans}}$ by considering the `crossing disk' model from \citet{cap13}, which consists of a circular outflow moving across a circular continuum source relative to our LOS (see their fig.~14; see also section 4.1 of \citealt{rog16}). In this model $v_{\rm{trans}}$ is expressed as $\sqrt{\Delta A} \ D_{\rm{cont}}/\Delta t$, where $\Delta A$ is the normalized flux difference between two spectra, $D_{\rm{cont}}$ is the diameter of the continuum source, and $\Delta t$ is the measured variability timescale between two spectra. 

We determine $D_{\rm{cont}}$ using the temperature function from standard accretion disk theory (see equation 3.20 from \citealt{pet97}) in combination with Wien's law (see also \citealt{ede15}). Determination of $D_{\rm{cont}}$ depends on the Eddington ratio $L_{\rm{bol}}/L_{\rm{Edd}}$, the SMBH mass $M_{\bullet}$, the radiative efficiency $\eta$, and the wavelength  $\lambda_{\rm{c}}$ of the accretion-disc radiation that is viewed by the BAL outflows. We calculate average values for log\,($L_{\rm{bol}}/L_{\rm{Edd}}$) and log\,$M_{\bullet}$ to be --0.3 and 9.5, respectively, using measurements taken from \citet{she11} for BAL quasars from our pristine sample (see Table~\ref{tab:pro}). We adopt a nominal value of $\eta=0.1$ (see e.g., \citealt{pet97}), and measure an average value of $\lambda_{\rm{c}}=1440$~\AA \ using our pristine disappearing and emerging BALs. The above measurements yield a continuum-source diameter on the order of $D_{\rm{cont}}=0.01$~pc for our pristine sample. 

The $D_{\rm{cont}}$ estimate, along with average measurements of $\Delta A=0.28$ and $\Delta t\le3.12$~yr (see Table~\ref{tab:mea}), allow a calculation of a nominal transverse outflow speed lower limit of \mbox{$v_{\rm{trans}} \ga 5400$~km~s$^{-1}$} for our pristine disappearing and emerging BALs. This crossing speed, combined with the average estimate for $M_{\bullet}$, yields a nominal upper limit for the outflow distance from the SMBH to be $r\la0.5$~pc, assuming Keplerian rotation and the crossing-disc geometry. Using the appropriate limiting measurements from our pristine sample, we obtain nominal minimum and maximum outflow distance upper limits to be on the order of $\la$ 0.01~pc and 10~pc, respectively. We measure an average time of $\Delta t'\le4.23$~yr for a complete BAL re-disappearance event to occur in our sample (see Fig.~\ref{fig:per}), which yields a nominal value for the transverse outflow size to be on the order of $D_{\rm{trans}}=0.01$~pc. Limiting measurements from our sample yield a minimum and maximum nominal transverse outflow size on the order of 0.001 and 0.1~pc, respectively. Our transverse outflow size estimate is of the same magnitude as our continuum-source size estimate, and is much larger than the radial outflow size of $D_{\rm{rad}}=1\times10^{-7}$~pc estimated in Section 5.2 using the ionization-change scenario.\footnote{We estimate a \emph{transverse} outflow size using the transverse-motion scenario, and obtain a \emph{radial} outflow size using the ionization-change scenario. It therefore remains unclear whether absorbers that cross our LOS exhibit distinct geometries from absorbers that respond to ionization fluctuations.}

Our order-of-magnitude estimates of the location and transverse size of disappearing/emerging BAL outflows increases the information available on quasar winds, as previous studies have primarily focused on constraining non-disappearing/non-emerging BAL flows. Our range of nominal BAL-outflow distances of \mbox{$r\la0.01-10$}~pc is consistent with theoretical models predicting that BAL outflows are launched from the accretion disc at sub-parsec distances from the SMBH (see e.g., \citealt{mur95}; \citealt{dek95}; \citealt{pro00}; \citealt{pro04}). We constrain BAL-outflow distances using BAL disappearance and emergence events, and our results are consistent with previous non-disappearing/non-emerging BAL-variability studies that attribute BAL changes to absorbers traversing our LOS. \citet{cap11,cap12,cap13} used  short-term and long-term variable C\thinspace\textsc{iv}/Si\thinspace\textsc{iv} BAL time-scales to constrain outflows to exist between 0.001 and 0.02~pc and $\sim$10~pc from the SMBH, respectively. \citet{cap14} placed an upper limit of 3.5~pc on the outflow distance using variable P\thinspace\textsc{v}, Si\thinspace\textsc{iv}, and C\thinspace\textsc{iv} BALs. Studies using variable Fe\thinspace\textsc{ii} and Mg\thinspace\textsc{ii} BALs constrained outflows to exist within tens of parsecs of the central source (\citealt{hal11}; \citealt{mcg15}). Previous studies using photoionization models have also constrained some BAL outflows to exist from parsecs to tens of parsecs from the SMBH using transitions from Fe\thinspace\textsc{ii}, Cr\thinspace\textsc{ii}, Mg\thinspace\textsc{ii}, and Mn\thinspace\textsc{ii} (\citealt{eve02}; \citealt{dek02}). 

Our nominal transverse BAL-outflow size estimate of $D_{\rm{trans}}=0.01$~pc favors models involving outflows in compact, dense geometries perhaps confined by magnetic fields (see e.g., \citealt{fuk10}). \citet{ham13} utilized an independent methodology from our work to investigate the absorption-line variability and X-ray absorption properties of 8 mini-BAL quasars, and concluded that the ionization levels of outflows in their sample are not moderated by the X-ray absorber (the so-called `radiative shield') but instead by high gas densities in small clouds (i.e., with radial outflow sizes $\la$~$1\times10^{-4}$~pc and transverse outflow sizes $\ga$~0.003~pc) (see also \citealt{gib09a}; \citealt{wu10}). The transverse-size lower limit for mini-BAL outflows from \citet{ham13} is consistent with our nominal transverse-size estimate for BAL outflows in our sample, suggesting that at least some mini-BALs and BALs may probe the same outflow system and both might exhibit dense, substructure geometries (see also the discussion at the end of Section 5.2).

\section{Conclusions}

We have conducted a quantitative, systematic investigation of C\thinspace\textsc{iv} and Si\thinspace\textsc{iv} BAL disappearance and emergence using a sample of 470 BAL quasars over \mbox{$\le$ 0.10--5.25}~yr in the rest frame with at least three spectroscopic epochs for each quasar from the SDSS, BOSS, and TDSS surveys. Below we summarize the most important findings from this analysis.

\begin{enumerate}

\item We detect 14 BAL quasars that exhibit highly significant C\thinspace\textsc{iv} and/or Si\thinspace\textsc{iv} BAL disappearance, and find 18 BAL quasars showing significant C\thinspace\textsc{iv} or Si\thinspace\textsc{iv} BAL emergence. Our sample consists of only BAL quasars. BAL disappearance and emergence in our sample occur over  \mbox{$\le1.46$--4.62}~yr, are rare events within our 470 BAL-quasar sample, exhibit weak line strengths compared to our full BAL sample, and show interesting EW and fractional EW changes relative to BAL variations more generally. The frequency of BAL to non-BAL quasar transition is 1.7$^{+0.8}_{-0.6}$ per cent given our dataset (see Section 4.1).

\item We detect four BALs that \emph{re-emerge} and three BALs that \emph{re-disappear} from C\thinspace\textsc{iv} or Si\thinspace\textsc{iv} in the available later TDSS observation using the significant cases of BAL disappearance and emergence in our sample (see Section 4.2 and 4.3). We also utilize BALs that re-emerge and BALs that do not re-emerge to obtain estimates for the BAL lifetime along our LOS to nominally be $\la$ 100--1000~yr for most BALs in our sample (see Section 5.1.1).

\item The behavior of our detected (re-)emerging and \mbox{(re-)disappearing} BALs collectively provide evidence supporting both transverse-motion and ionization-change scenarios to explain BAL disappearance and emergence. Evidence of saturation in some detected BALs supports models involving transverse motions (see Section 5.1.3), while observed evidence for ionization changes include preservation of C\thinspace\textsc{iv} BAL profiles through Si\thinspace\textsc{iv} re-emergence and re-disappearance at the same velocity, coordinated/coherent BAL variations over a large velocity interval, and significant BEL and continuum fluctuations (see Section 5.1.2).

\item Under the ionization-change scenario, the disappearing and emerging BAL outflows we probe would exist at a nominal location on the order of $\la$100~pc from the SMBH and exhibit a typical radial size on the order of $\ga1\times10^{-7}$~pc. Under the transverse-motion scenario, the disappearing and emerging BAL outflows in our sample would exist at a nominal distance of $\la$0.5~pc from the central source and exhibit a nominal transverse size on the order of $\sim$0.01~pc. These estimates are broadly consistent with theoretical models and previous observational studies of non-disappearing/non-emerging quasar winds, and provide evidence supporting compact outflow geometries (see Section 5.2 and 5.3).

\end{enumerate}

Utilizing observations from the SDSS, BOSS, and TDSS surveys has allowed us to detect and characterize systematically BAL re-disappearance and re-emergence events, which would not have been possible with previous, small-sized, two-epoch datasets. The ongoing acquisition of TDSS data will allow future, large-sample studies to characterize further the behavior during and after BAL disappearance and emergence, which will provide tighter constraints on BAL-outflow lifetimes and geometries to inform theoretical models of quasar winds. The initial sample of 2005 BAL quasars has thus far enabled large-scale, systematic investigations into BAL variability over multi-year time-scales (see \citealt{fil12,fil13,fil14}; \citealt{gri16}), and will allow for numerous future studies to advance our understanding of quasar winds. For example, this large sample will enable a systematic investigation between BAL variations and BEL changes/properties from ions including C\thinspace\textsc{iv}, Si\thinspace\textsc{iv}, Al\thinspace\textsc{iii}, C\thinspace\textsc{iii}], and Mg\thinspace\textsc{ii}. It will also be interesting to conduct a comparative study between variable and non-variable BALs from C\thinspace\textsc{iv} and Si\thinspace\textsc{iv} in order to understand better the apparent stability of a large percentage of quasar outflows. 

\section*{Acknowledgements}

We thank the referee for a constructive report. We acknowledge support from NSF grant AST--1516784 (SMM, WNB), NSF grant AST--1517113 (WNB, CJG, DPS), NSERC (PBH), and TUBITAK 115F037 (NFA).

Funding for the Sloan Digital Sky Survey IV has been provided by the Alfred P. Sloan Foundation, the U.S. Department of Energy Office of Science, and the Participating Institutions. SDSS-IV acknowledges support and resources from the Center for High-Performance Computing at the University of Utah. The SDSS web site is www.sdss.org.

SDSS-IV is managed by the Astrophysical Research Consortium for the Participating Institutions of the SDSS Collaboration including the Brazilian Participation Group, the Carnegie Institution for Science, Carnegie Mellon University, the Chilean Participation Group, the French Participation Group, Harvard-Smithsonian Center for Astrophysics, Instituto de Astrof\'isica de Canarias, The Johns Hopkins University, Kavli Institute for the Physics and Mathematics of the Universe (IPMU) / University of Tokyo, Lawrence Berkeley National Laboratory, Leibniz Institut f\"ur Astrophysik Potsdam (AIP),  Max-Planck-Institut f\"ur Astronomie (MPIA Heidelberg), Max-Planck-Institut f\"ur Astrophysik (MPA Garching), Max-Planck-Institut f\"ur Extraterrestrische Physik (MPE), National Astronomical Observatories of China, New Mexico State University, New York University, University of Notre Dame, Observat\'ario Nacional / MCTI, The Ohio State University, Pennsylvania State University, Shanghai Astronomical Observatory, United Kingdom Participation Group, Universidad Nacional Aut\'onoma de M\'exico, University of Arizona, University of Colorado Boulder, University of Oxford, University of Portsmouth, University of Utah, University of Virginia, University of Washington, University of Wisconsin, Vanderbilt University, and Yale University.

The CSS survey is funded by the National Aeronautics and Space Administration under Grant No. NNG05GF22G issued through the Science Mission Directorate Near-Earth Objects Observations Program.  The CRTS survey is supported by the U.S.~National Science Foundation under grants AST-0909182 and AST-1313422.








\appendix

\section{Notes on Individual Objects}

Below we briefly report on BAL re-emergence/re-disappearance events, variability cases of special interest, and problematic flux offsets that are somewhat ambiguous in their origin. We consider the cases listed here as being representative of the range of variations/offsets that we observe across our entire pristine sample.

\emph{J020629.32+004843.0} exhibits a pristine case of Si\thinspace\textsc{iv} BAL re-disapperance centered at --9000~km~s$^{-1}$ (see Fig.~\ref{fig:per}). There appears to be a small amount of potential residual absorption at --11\,000~km~s$^{-1}$ before the BAL emerges (black spectrum) and after the BAL disappears (blue spectrum), however this feature can be considered potentially spurious given the amplitude of the noise on either side of the BAL. The C\thinspace\textsc{iv} BAL varies at the same velocity as the re-disappearing Si\thinspace\textsc{iv} (i.e., at 1500\,\AA; see Fig.~\ref{fig:var}), and we also observe C\thinspace\textsc{iv} BAL variability near 1470\,\AA \ and C\thinspace\textsc{iv} BEL changes near 1540\,\AA \ in this quasar (see Fig.~\ref{fig:var}).

\emph{J023252.80-001351.1} exhibits a pristine case of Si\thinspace\textsc{iv} BAL re-emergence centered at --7900~km~s$^{-1}$ (see Fig.~\ref{fig:pdr}). There appears potentially to be small amounts of residual absorption after the BAL disappears (see red spectrum); however, these features are likely spurious given the amplitude of the noise on either side of the BAL. It is also noteworthy that the re-emerging trough (blue spectrum) appears to kinematically match the profile of the BAL before it disappeared (black spectrum); there are matching absorption features at --6000 and --9000~km~s$^{-1}$. The C\thinspace\textsc{iv} BAL decreases in strength at the same velocity as the disappearing Si\thinspace\textsc{iv} BAL (i.e., at 1500\,\AA; see Fig.~\ref{fig:var}), and we observe C\thinspace\textsc{iv} BAL variations near 1525\,\AA \ as well as Si\thinspace\textsc{iv} and C\thinspace\textsc{iv} BEL changes near 1400 and 1550\,\AA, respectively, in this source (see Fig.~\ref{fig:var}).

\emph{J091944.53+560243.3} shows a pristine case of C\thinspace\textsc{iv} BAL re-emergence centered at --14\,000~km~s$^{-1}$ (see Fig.~\ref{fig:pdr}). It is interesting that only the high-velocity portion of the BAL re-emerged. We also observe emerging C\thinspace\textsc{iv} absorption from \mbox{--15\,000} to \mbox{--18\,000}~km~s$^{-1}$ (blue spectrum) during the C\thinspace\textsc{iv} BAL re-emergence event. These observations indicate potentially complex BAL outflow geometries or a complicated combination of ionization changes and transverse motions occurring in these absorbers. We also observe a decrease in the C\thinspace\textsc{iv} BEL strength near 1550\,\AA \ during the C\thinspace\textsc{iv} BAL disappearance event in this object (see Fig.~\ref{fig:var}).

\emph{J092434.54+423615.1} shows a pristine case of C\thinspace\textsc{iv} BAL re-disappearance centered at --9500~km~s$^{-1}$ (see Fig.~\ref{fig:per}). There appears to be small amounts of potential residual absorption after the BAL re-disappears (blue spectrum) at \mbox{--9000} and --11\,000~km~s$^{-1}$, although the latter feature is located just outside the region where the BAL exists. If these features are true absorption and not spurious, it is ambiguous whether the BAL in the blue spectrum is in the process of re-disappearing or has already re-disappeared and is in the process of re-emerging. We also observe significant C\thinspace\textsc{iv} BAL variations between --5000 and 0\,km\,s$^{-1}$ during the C\thinspace\textsc{iv} BAL re-disappearance event in this quasar (see Fig.~\ref{fig:per}).

\emph{J093243.25+414230.8} exhibits a pristine case of a C\thinspace\textsc{iv} BAL re-disapperance event centered at --22\,500~km~s$^{-1}$ (see Fig.~\ref{fig:per}). The BAL variation during the emergence event (between the black and red spectra) occurs over many consecutive pixels, and the observed variation is therefore likely real and not spurious. The C\thinspace\textsc{iv} BAL emergence event is accompanied by decreases in strength from the Si\thinspace\textsc{iv} BEL, and these two events collectively cause decreases in normalized flux from --30\,000 to \mbox{--20\,000}\,km\,s$^{-1}$ (see Fig.~\ref{fig:per}). We also observe variations in the C\thinspace\textsc{iv} BAL near --14\,000\,km\,s$^{-1}$ (see Fig.~\ref{fig:per}) as well as decreases in the C\thinspace\textsc{iv} BEL line strength near 1550\,\AA \ (see Fig.~\ref{fig:var}).

\emph{J094456.75+544117.9} exhibits a pristine case of C\thinspace\textsc{iv} BAL re-emergence centered at --14\,000~km~s$^{-1}$ (see Fig.~\ref{fig:pdr}). The trough re-emerges at the same velocity (blue spectrum) as the original BAL (black spectrum), and the profiles for both troughs are relatively smooth and exhibit similar shapes. The apparent variations at $<$1375\,\AA \ during the C\thinspace\textsc{iv} BAL disappearance event are not real, but are due to difficulty in normalizing the continuum over this wavelength interval (see Fig.~\ref{fig:var}).

\emph{J100249.66+393211.0} shows a pristine case of Si\thinspace\textsc{iv} BAL re-emergence centered at --7000~km~s$^{-1}$ (see Fig.~\ref{fig:pdr}). The re-emerging trough (blue spectrum) lies at the same velocity as the original BAL (black spectrum), and both troughs exhibit kinematic features that are similar to one another (i.e., both exhibit multiple absorption substructures within their profiles). We observe significant C\thinspace\textsc{iv} BAL variations at the same velocity as the Si\thinspace\textsc{iv} BAL re-emergence event in this object (i.e., near 1520\,\AA; see Fig.~\ref{fig:var}).

\emph{J100318.99+521506.3} exhibits a pristine case of Si\thinspace\textsc{iv} BAL emergence centered at 1325\,\AA \ (see Fig.~\ref{fig:var}). We observe flux offsets in both comparisons in Fig.~\ref{fig:var} at $>$1525\,\AA, which are likely attributed to either a complex combination of broad emission changes or difficulties in normalizing the continua.

\emph{J100906.61+490127.7} shows a pristine case of C\thinspace\textsc{iv} BAL emergence centered at $\sim$1470\,\AA \ (see Fig.~\ref{fig:var}). The comparison after the BAL emergence event occurs [i.e., the red spectrum (MJD 56366) compared to the blue spectrum (MJD 56660) in Fig.~\ref{fig:var}] exhibits a continuous flux offset at $<$1450\,\AA. This offset is probably due to difficulties in normalizing the continuum for these spectra.

\emph{J111342.27+580407.6} exhibits a pristine case of C\thinspace\textsc{iv} BAL disappearance centered at 1375\,\AA \ (see Fig.~\ref{fig:var}). The spectrum comparison after the disappearance event [i.e., the red spectrum (MJD 56666) compared to the blue spectrum (MJD 57374) in Fig.~\ref{fig:var}] shows flux offsets between \mbox{1400--1500}\,\AA \ and at $<1375$\,\AA. These regions may be due to a complex combination of BAL variations and broad emission changes, and/or due to difficulties in normalizing the continua.


\bsp	
\label{lastpage}
\end{document}